\documentclass[12pt]{article} 
\usepackage[sectionbib]{natbib}
\usepackage[]{hyperref}  
\usepackage{amsmath}
\usepackage{amssymb}
\usepackage{amsfonts}
\usepackage{multirow}
\usepackage{amsthm}
\usepackage[utf8]{inputenc}
\usepackage{graphicx, subfig}
\usepackage{algorithmic, algorithm}
\usepackage{tikz}
\usetikzlibrary{trees}
\usepackage{bm}

\def\version{0}
\if\version0
        \usepackage[margin = 2.5cm]{geometry}
        \hypersetup{
  colorlinks   = true, 
  urlcolor     = {green!50!black}, 
  linkcolor    = {green!50!black}, 
  citecolor   = {green!50!black} 
}
\else
	\usepackage{array,epsfig,fancyheadings,rotating}
	\usepackage{sectsty, secdot}
	\sectionfont{\fontsize{12}{14pt plus.8pt minus .6pt}\selectfont}
	\renewcommand{\theequation}{\thesection\arabic{equation}}
	\subsectionfont{\fontsize{12}{14pt plus.8pt minus .6pt}\selectfont}
	\usepackage{xr}
\fi

\theoremstyle{definition}

\newcommand{\real}{\mathbb{R}}
\newcommand{\N}{\mathcal{N}}
\DeclareMathOperator{\mse}{MSE}
\DeclareMathOperator{\rmse}{RMSE}
\def\rmd{\mathrm{d}}
\newcommand{\ess}{\textrm{ESS}}
\newcommand{\cess}{\textrm{CESS}}
\newcommand{\ks}{\textrm{KS}}

\newcommand{\Id}{\textrm{Id}}

\usepackage[textwidth=1.8cm, textsize=scriptsize]{todonotes}
\usepackage{setspace} 

\title{A divide and conquer sequential Monte Carlo approach to high dimensional filtering}
\author{
  Francesca R. Crucinio\thanks{Department of Statistics, University of Warwick, Coventry, CV4 7AL, UK. Email: $\{$francesca.crucinio, a.m.johansen$\}$@warwick.ac.uk} 
  \and
  Adam M. Johansen\footnotemark[1] 
}

\date{}

\graphicspath{{Images/}}

\begin{document}
	
\if\version1 
	\renewcommand{\baselinestretch}{2}
	
	\markright{ \hbox{\footnotesize\rm Statistica Sinica
		}\hfill\\[-13pt]
		\hbox{\footnotesize\rm
		}\hfill }
	
	\markboth{\hfill{\footnotesize\rm FRANCESCA R. CRUCINIO AND ADAM M. JOHANSEN} \hfill}	
	{\hfill {\footnotesize\rm DIVIDE-AND-CONQUER PARTICLE FILTERING} \hfill}
	
	\renewcommand{\thefootnote}{}
	$\ $\par
	
	
	\fontsize{12}{14pt plus.8pt minus .6pt}\selectfont \vspace{0.8pc}
	\centerline{\large\bf A DIVIDE-AND-CONQUER SEQUENTIAL MONTE CARLO}
	\vspace{2pt} 
	\centerline{\large\bf APPROACH TO HIGH DIMENSIONAL FILTERING}
	\vspace{.4cm} 
	\centerline{Francesca R. Crucinio and Adam M. Johansen} 
	\vspace{.4cm} 
	\centerline{\it Department of Statistics, University of Warwick}
	\vspace{.55cm} \fontsize{9}{11.5pt plus.8pt minus.6pt}\selectfont
	
	
	\begin{quotation}
		\noindent {\it Abstract:}
		We propose a divide-and-conquer approach to filtering which decomposes the state variable into low-dimensional components to which standard particle filtering tools can be successfully applied and recursively merges them to recover the full filtering distribution. It is less dependent upon factorization of transition densities and observation likelihoods than competing approaches and can be applied to a broader class of models. Performance is compared with state-of-the-art methods on a benchmark problem and it is demonstrated that the proposed method is broadly comparable in settings in which those methods are applicable, and that it can be applied in settings in which they cannot.
		
		\vspace{9pt}
		\par
	\end{quotation}\par

	\def\thefigure{\arabic{figure}}
	\def\thetable{\arabic{table}}
	
	\renewcommand{\theequation}{\thesection.\arabic{equation}}

	\fontsize{12}{14pt plus.8pt minus .6pt}\selectfont

\else
\maketitle
\begin{abstract}
		
				We propose a divide-and-conquer approach to filtering which decomposes the state variable into low-dimensional components to which standard particle filtering tools can be successfully applied and recursively merges them to recover the full filtering distribution. It is less dependent upon factorization of transition densities and observation likelihoods than competing approaches and can be applied to a broader class of models. Performance is compared with state-of-the-art methods on a benchmark problem and it is demonstrated that the proposed method is broadly comparable in settings in which those methods are applicable, and that it can be applied in settings in which they cannot.
		\end{abstract}
\fi

\section{Introduction}
Particle filters (PFs), an instance of sequential Monte Carlo (SMC) methods, are a popular class of algorithms to perform state estimation for state space models (SSM) --- or general-state-space hidden Markov models as they are sometimes known. We consider the class of SSMs with a latent $\real^d$-valued process $( X_t)_{t\geq 1}$ and conditionally independent $\real^p$-valued observations $(Y_t)_{t\geq 1}$. Such a SSM $( X_t, Y_t)_{t\geq 1}$ is defined by the transition density $f_t(x_{t-1},x_t)$ of the latent process, with the convention that $f_1(x_{0}, x_1)\equiv f_1(x_1)$, and the observation likelihood $g_t(y_t|x_t)$.
In this work, we are interested in approximating the sequence of filtering distributions, $(p(x_t | y_{1:t}))_{t\geq 1}$, i.e. at each time $t$ the distribution of the latent state at that time given the observations obtained by that time.

Basic PF algorithms are known to suffer from the curse of dimensionality, requiring an exponential increase in computational requirements as the dimension $d$ grows, limiting their applicability to large systems \citep{rebeschini2015can, bengtsson2008curse}. While the ensemble Kalman filter \citep{evensen2009data} can tackle high dimensional problems, it involves approximations which do not disappear even asymptotically and does not perform well if the model is far from linear and Gaussian \citep{lei2010comparison}.

To extend the use of particle filters to high dimensional problems it is natural to attempt to exploit the fact that dependencies in high dimensional SSMs encountered in practice are often local in space in order to decompose the filtering problem into a collection of local low-dimensional problems which can somehow be combined; examples of this strategy are the block particle filter \citep{rebeschini2015can}, space-time particle filters (STPF; \cite{beskos2017stable}) and nested sequential Monte Carlo (NSMC) methods \citep{naesseth2015nested, naesseth2019high}.

We propose a divide-and-conquer approach in which the state space is divided into smaller subsets over which standard particle filtering ideas can be applied, these smaller subsets are then recursively merged in a principled manner to obtain approximations over the full state space.
Our method is an extension of the divide-and-conquer sequential Monte Carlo (DaC-SMC) algorithm introduced by \cite{lindsten2017divide} to the filtering context, which exploits ideas akin those in \cite{klaas2012toward, lin2005independent} to marginalize out the past $x_{1:t-1}$ at a given time $t$.

In order to apply DaC-SMC to the filtering problem, we define a non-standard sequence of targets evolving both in space and in time: at a given time $t$ we define $d$ univariate targets serving as proxies for the marginals of the filtering distribution, $p(x_t(i)\vert y_{1:t})$ for $i=1,\dots, d$, we then iteratively combine these lower dimensional targets to obtain approximations of higher-dimensional marginals (e.g. $p(x_t(i:i+1)\vert y_{1:t})$)  until we recover the full filtering distribution, $p(x_t\vert y_{1:t}) \equiv p(x_t(1:d)\vert y_{1:t})$.

Unlike NSMC and STPF, this approach does not require analytic expressions for marginals of the transition density or observation likelihood, but only point-wise evaluations of $f_t$ and $g_t$, making it suitable to tackle high dimensional SSM in which the observations are correlated in non-trivial ways (cf. the model in Section~\ref{sec:ng}), as is common in real applications (see, e.g., \citet[Section 2]{chib2009multivariate}). 

We review the basic ideas of particle filtering, its marginal variant and divide-and-conquer SMC in Section~\ref{sec:background}; we then show how to extend the ideas underlying DaC-SMC to the filtering problem in Section~\ref{sec:dac_filter}, where we also discuss strategies to improve computational cost and accuracy. Finally, in Section~\ref{sec:expe} we compare the performances of the divide-and-conquer approach with  NSMC and STPF on a simple linear Gaussian SSM for which the Kalman filter provides the exact filtering distribution.
Our experiments show that the errors in approximating the true filtering distribution obtained with the divide-and-conquer approach are comparable to those of NSMC and STPF.
We then consider a spatial model whose correlation structure in the observation model makes it impossible to apply NSMC or STPFs (at least without additional approximations) and empirically show that the proposed approach can recover stable estimates of the filtering distribution which, in small dimensional settings, coincide with those obtained with a bootstrap PF with a large number of particles.

\section{Background}
\label{sec:background}
\subsection{Particle Filtering}

We describe here the basic SMC approach, often referred to as sequential importance resampling \cite[p.\ 15]{doucet2011}, and refer to \cite{liu2001monte,chopin2020} for a more extensive treatment.

Given the sequence of unnormalized target densities $(\gamma_t)_{t\geq 1}$, with
\begin{align}
\label{eq:joint}
\gamma_t(x_{1:t}) = p(x_{1:t}, y_{1:t}) =\prod_{k=1}^t f_k(x_{k-1}, x_k)g_k(y_k\vert x_k),
\end{align}
defined on $(\real^d)^t$, PFs proceed iteratively, and, at time $t-1$ approximate $$\pi_{t-1}:=\gamma_{t-1}/\int \gamma_{t-1}(x_{1:t-1}) \rmd x_{1:t-1}$$ with a cloud of  particles $\{x_{1:t-1}^{n}\}_{n=1}^N$. The particles are propagated forward in time using a Markov kernel $K_t(x_{1:t-1}, \cdot)$, reweighted using the weight function $w_t:=\gamma_t/\gamma_{t-1}\otimes K_t$ and resampled to obtain a new particle population $\{x_{1:t}^{n}\}_{n=1}^N$ approximating $\pi_t$.

Standard PFs formally target distributions~\eqref{eq:joint} whose dimension increases at every time step $t$, although one is often interested only in the final time marginal of the distributions so approximated, in this case the filtering distribution. An alternative to this approach is given by marginal particle filters (MPFs; \cite{klaas2012toward}) and the closely related ideas of \citet{lin2005independent}. MPFs target the filtering distribution directly
\begin{align}
\label{eq:joint_marginal}
\gamma_t(x_{t}) = p(x_{t}\vert y_{1:t}) =g_t(y_t\vert x_t)\int f_t(x_{t-1}, x_t)p(x_{t-1}\vert y_{1:t-1}) \rmd x_{t-1};
\end{align}
and, since the integral w.r.t. $x_{t-1}$ is intractable, replace $p(x_{t-1}\vert y_{1:t-1})$ with its particle approximation obtained at time $t-1$.
Given the new sequence of targets, MPFs proceed as standard PFs, with the only difference being that, whenever we need to compute an integral w.r.t. $x_{t-1}$, this is approximated using $\pi_{t-1}^N$, obtained by normalizing $\gamma_{t-1}^N$, the particle approximation of $p(x_{t-1}\vert y_{1:t-1})$.
Basic MPFs incur an $O(N^2)$ cost for each time step, because of the presence of the integral w.r.t. $x_{t-1}$ in the weight computations, although lower cost strategies might be employed in some cases (\citet{lin2005independent}; see also \cite{klaas2006fast}).

\subsection{Particle Filters for High Dimensional Problems}
We briefly summarize three classes of particle filters which make use of space decompositions to tackle the filtering problem which we believe to be the state-of-the-art in Monte Carlo approximation of high dimensional filtering distributions.

The block particle filter (BPF; \cite{rebeschini2015can}) algorithm relies on a decomposition of the state space $\real^d$ into lower dimensional blocks on which, at each $t$, one step of a standard PF is run. The approximation of the filtering distribution over the whole state space is obtained by taking the product of the lower dimensional approximations on each block. BPFs are inherently biased because of the decomposition into blocks; although this bias can be eliminated asymptotically by allowing the blocks to grow at an appropriate rate with computational effort.

Space-time particle filters (STPF; \cite{beskos2017stable}) exploit local dependence structures in the observation $y_t$ to gradually introduce the likelihood term by decomposing the space dimension into smaller subsets and running $N$ independent particle filters on each of the subsets (also called islands) which are then combined using an importance resampling step, the presence of the latter, guarantees asymptotically consistent approximations, contrary to BPFs.
Crucial to the implementation of STPF is that the joint law~\eqref{eq:joint} at time $t$ can be factorized so that the marginal of $x_{t}(i)$ given the observations and the past only depends on a neighbourhood of $x_t(i)$, $\{x_t(j):\ j\in \mathcal{A}\}$ for some $\mathcal{A}\subset \{1,\dots, d\}$, for all $i=1, \dots, d$. STPFs are particularly amenable to SSM which are time discretizations of SDEs, since in this case one can build time discretization schemes which guarantee analytical forms for the marginals \citep{akyildiz2022space}.
A marginal version of STPFs also exists \citep{beskos2017stable, xu2019particle}.

Nested sequential Monte Carlo (NSMC; \cite{naesseth2015nested}) treats the problem of recovering the filtering distribution as a smoothing problem, where the time variable is replaced by the dimension $d$, approximates the fully adapted proposal of \citet{pitt1999filtering} with an inner SMC iteration and then uses the result in the outer level which corresponds to a standard forward filtering backward simulation algorithm. 
NSMC is particularly well-suited for Markov random fields in which the temporal and the spatial components can be separated, as this makes the backward simulation straightforward.

\subsection{Divide and Conquer SMC}
\label{sec:dac}
Divide-and-Conquer SMC (DaC-SMC; \cite{lindsten2017divide}) is an extension of standard SMC in which a collection of (unnormalized) target distributions $( \gamma_u)_{u\in\mathbb{T}}$ is indexed by the nodes of a rooted tree, $\mathbb{T}$, and particles evolve from the leaves to the root, $\mathfrak{R}$, rather than along a sequence of distributions indexed by (algorithmic) time. It shares many of the convergence properties of standard SMC \citep{Kuntz2021}.

The target distributions are defined on spaces whose dimension grows as we progress up the tree: for each $u$, $\pi_u \propto \gamma_u$ is a density over $\real^{\vert \mathbb{T}_u \vert}$ where
$\mathbb{T}_u$ denotes the sub-tree of $\mathbb{T}$ rooted at $u$ (obtained by removing all nodes from $\mathbb{T}$ except for $u$ and its descendants) and $\vert \mathbb{T}_u\vert$ denotes its cardinality.
We focus here on the case in which the state space is $\real^d$, however, essentially the same construction allows for much more general spaces, including those with discrete components.

As in standard SMC, each distribution $\gamma_u$ is approximated by a particle population $\{x_u^{n}\}_{n=1}^N$. However, these distributions do not evolve `linearly' but are merged whenever the corresponding branches of $\mathbb{T}$ merge. For simplicity, we describe here the case in which $\mathbb{T}$ is a binary tree and each non-leaf node $u$ has two children, a left child $\ell(u)$ and a right child $r(u)$.

If $u$ is a leaf node, the algorithm performs a simple importance sampling step with proposal $K_u$ and importance weight $w_u := \gamma_u/K_u$ to obtain a weighted particle population $\{x_u^{n}, w_u^{n}\}_{n=1}^N$ approximating $\gamma_u$.
Otherwise, to obtain a particle population approximating $\gamma_u$, we gather the particle populations associated with each of $u$'s children and compute the (weighted) product form estimator \citep{Kuntz2021product}
\begin{align}
\label{eq:pf}
\gamma_{\mathcal{C}_u}^N = \frac{1}{N^2}\sum_{n_1=1}^N\sum_{n_2=1}^N w_{\ell(u)}(x_{\ell(u)}^{n_1})w_{r(u)}(x_{r(u)}^{n_2}) \delta_{(x_{\ell(u)}^{n_1}, x_{r(u)}^{n_2})}
\end{align}
to approximate the product of the marginal distributions $\gamma_{\mathcal{C}_u}:= \gamma_{\ell(u)}\times \gamma_{r(u)}$.
The $O(N^2)$ cost of evaluating $\gamma_{\mathcal{C}_u}^N$ can be prohibitively large, we discuss lower cost alternatives in Section~\ref{sec:light}.

We reweight the particle approximation $\gamma_{\mathcal{C}_u}^N$ of $\gamma_{\mathcal{C}_u}$ to target $\gamma_u$; the resulting \emph{mixture} (importance) weights 
\begin{align}
\label{eq:w_mix_dac}
m_u(x_{\ell(u)}, x_{r(u)}):= \frac{\gamma_u(x_{\ell(u)}, x_{r(u)})}{\gamma_{\ell(u)}(x_{\ell(u)})\gamma_{r(u)}(x_{r(u)})}
\end{align}
capture the mismatch between $\gamma_u$ and $\gamma_{\ell(u)} \times \gamma_{r(u)}$, and are incorporated prior to resampling similarly to the auxiliary ``twisting'' function in the auxiliary PF (see, e.g., \citet[Chapter 10]{chopin2020}). 
This leads to weights of the form
\begin{align}
\label{eq:updated_weights_dac}
\tilde{w}_u(x_{\ell(u)}, x_{r(u)}):= w_{\ell(u)}(x_{\ell(u)})w_{r(u)}(x_{r(u)})m_u(x_{\ell(u)}, x_{r(u)}).
\end{align}
Resampling $N$ times from $\tilde{w}_{u}\gamma_{\mathcal{C}_u}^N$, using any unbiased resampling scheme (cf. \citet{Gerber2019}), we obtain an equally weighted particle population $\{\tilde x_{u}^{n}, w_{u}^{n}=1\}_{n=1}^N$ approximating $\gamma_u$. If necessary, we can then apply a $\pi_u$-invariant Markov kernel $K_u$, to rejuvenate the particles. Algorithm~\ref{alg:dac}, which is applied to the root node to carry out the sampling process, summarizes this.

The DaC approach in Algorithm~\ref{alg:dac} is a special case of that considered in \citet{lindsten2017divide, Kuntz2021}, in which the target at each non-leaf node is defined on the product of the spaces on which each of its child targets are defined.
DaC-SMC is particularly amenable to distributed implementation \citep[Section 5.3]{lindsten2017divide}.

\begin{algorithm}[th]
\begin{algorithmic}[1]
\IF{$u$ is a leaf}
\STATE{\textit{Initialize:} draw $ x_{u}^{n}\sim K_{u}$ and compute $w_{u}^{n}=\gamma_u/K_u$ for all $n\leq N$.}
\ELSE
\STATE{\textit{Recurse:} set $(\{x_{v}^{n}, w_{v}^{n}\}_{n=1}^N):=\text{dac\_smc}(v)$ for $v$ in $\{\ell(u), r(u)\}$ and obtain $\gamma_{\mathcal{C}_u}^N$ in~\eqref{eq:pf}.}
\STATE{\textit{Merge:} compute $\tilde{w}_{u}^{(n_1, n_2)}$ in~\eqref{eq:updated_weights_dac} for all $n_1, n_2\leq N$.}
\STATE{\textit{Resample:} draw $\{\tilde x_{u}^{n}\}_{n=1}^N$ using weights $\tilde{w}_{u}^{(n_1, n_2)}$ and set $w_{u}^{n}=1$ for all $n\leq N$.}
\STATE{\textit{(Optionally):} draw $ x_{u}^{n}\sim K_{u}(\tilde x_{u}^{n},\cdot)$ for all $n\leq N$.}\newline
{\textit{(Otherwise):} set $x_u^n = \tilde{x}_u^n$ for all $n\leq N$.}
\ENDIF

 \end{algorithmic}
 \caption{dac\_smc$(u)$ for $u$ in $\mathbb{T}$.}\label{alg:dac}
\end{algorithm}

\section{Divide and Conquer within Marginal SMC for Filtering}
\label{sec:dac_filter}

To apply the DaC-SMC algorithm described above to the filtering problem, we need to identify a suitable collection $( \widetilde{\gamma}_{t,u})_{ u\in\mathbb{T}}$ indexed by the nodes of the tree $\mathbb{T}$, for each time $t$.
Graphically, this corresponds to a path graph (corresponding to time) in which each node has associated with it a copy of the tree $\mathbb{T}$ (corresponding to space).
In this case Algorithm~\ref{alg:dac} takes as input at the leaves a particle population approximating the filtering distribution at time $t-1$ and outputs at the root a particle population approximating the filtering distribution at time $t$.

At a given time $t$, to build the collection $( \widetilde{\gamma}_{t,u})_{u\in\mathbb{T}}$, we consider spatial decompositions of $x_t$ into low dimensional (often univariate) elements. Here, we consider a simple decomposition obtained  by identifying the $d$ components $(x_t(1), \dots, x_t(d))$ with the leaves of a tree $\mathbb{T}$. As we move up the tree, the components are merged pairwise until $x_t=x_t(1:d)$ is recovered at the root node $\mathfrak{R}$.
For simplicity, we assume that $d=2^D$ for some $D\in\mathbb{N}$ so that $\mathbb{T}$ is a perfect binary tree; essentially the same construction applies to general $d$.

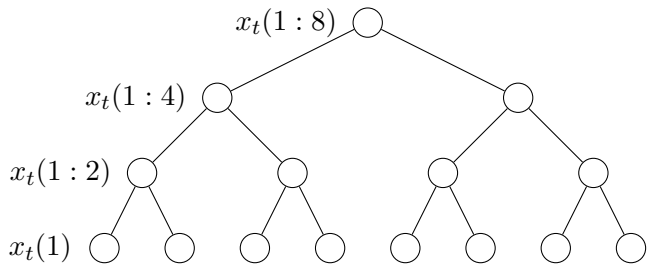
\begin{figure}
\centering
\begin{tikzpicture}[level distance=1cm,
level 1/.style={sibling distance=4cm},
level 2/.style={sibling distance=2cm},
level 3/.style={sibling distance=1cm}]
   \tikzstyle{every node}=[circle, draw, font =\small]

\node (Root1)  [label={ left:{$x_t(1:8)$}}] {}
    child {
    node [label={ left:{$x_t(1:4)$}}] {} 
    child { node [label={ left:{$x_t(1:2)$}}] {}
    	child { node [label={ left:{$x_t(1)$}}] {} }
    	child { node {} }
        }
    child { node {} 
    	child { node {} }
    	child { node {} }
        }
}
child {
    node {}
    child { node {} 
    	child { node {} }
    	child { node {} }
       }
    child { node {} 
    	child { node {} }
    	child { node {} }
        }
};
\end{tikzpicture} 
\caption{Space decomposition for $d=8$.}
\label{fig:tree}
\end{figure}

We denote the set of components associated with node $u$ by $\mathcal{V}_u$, its cardinality increases from leaves to root: at the level of the leaves $\vert \mathcal{V}_u\vert =1$, while $\vert \mathcal{V}_{\mathfrak{R}}\vert =d$. Figure~\ref{fig:tree} shows the space decomposition for $d=8$.

As the filtering problem has an inherent (temporal) sequential structure, the collection $( \widetilde{\gamma}_{t,u})_{u\in\mathbb{T}}$ at time $t$ is most easily specified in terms of the filtering distribution at time $t-1$, $\widetilde{\gamma}_{t-1,\mathfrak{R}}$ --- as shown below in \eqref{eq:gamma}. Similarly to MPFs, we deal with this dependence by approximately marginalizing out the previous timestep using the existing sample approximation.
We introduce auxiliary functions
$f_{t,u}: \real^{d}\times\real^{\vert \mathcal{V}_u\vert} \to \real$ and $g_{t,u}:\real^{\vert \mathcal{V}_u\vert}\times\real^{\vert \mathcal{V}_u\vert}\to \real$ for $t\geq 1$ and $u\in\mathbb{T}$, such that $f_{t,\mathfrak{R}}=f_t, g_{t,\mathfrak{R}} = g_t$ and for $u \in \mathbb{T}\setminus\mathfrak{R}$, $f_{t,u}$ and $g_{t,u}$ serve as proxies for marginals of the transition density and observation likelihood, respectively.
These auxiliary functions are used to define our collection of target densities $(\widetilde{\gamma}_{t, u})_{t\geq 1, u\in\mathbb{T}}$ over $\real^{\vert \mathcal{V}_u \vert}$: 
\begin{align}
\label{eq:gamma}
   \widetilde{\gamma}_{t,u}(z_{t,u}) &= 
    g_{t,u}(z_{t,u}, (y_t(i))_{i\in\mathcal{V}_{u}}) \int f_{t,u}(x_{t-1}, z_{t,u})\widetilde{\gamma}_{t-1, \mathfrak{R}}(x_{t-1})\rmd x_{t-1}
\end{align}
where $z_{t,u}=(x_t(i))_{i\in\mathcal{V}_{u}}$ are the components of $x_t$ associated with node $u$ and $x_{t-1}$ denotes the previous state of the system.
The requirement that $g_{t,\mathfrak{R}} = g_t,  f_{t,\mathfrak{R}}=f_t$ ensures that at the root we obtain the distribution in~\eqref{eq:joint_marginal}. 
The integral w.r.t. $\widetilde{\gamma}_{t-1, \mathfrak{R}}$ in~\eqref{eq:gamma} cannot be computed analytically, hence, as in MPFs, we use a sample approximation of this integral, as described in the next section.

If the marginals of $f_t, g_t$ are available, one could use them to define $g_{t,u}, f_{t,u}$, but this is not essential: these intermediate distributions can be essentially arbitrary up to the absolute continuity required to justify the importance sampling steps although, of course, the variance of the estimator will be influenced by this choice. The issue of specifying these distributions is closely related to that of choosing the sequence of artificial targets in a standard SMC sampler when one is only interested in the final target distribution (see \cite{del2006sequential}, where they note that optimizing this sequence is a very difficult problem). See \citet[Section 4.2]{Kuntz2021} for a theoretical perspective in the divide-and-conquer context, which suggests that the optimal choice of intermediate target distribution is the appropriate marginal of the root target, hence the suggestion to use the marginals of $f_t$ and $g_t$ where these are available. In practice, when doing this exactly is not feasible, this perspective suggests that we should seek to approximate these distributions with distributions with comparable tail behaviour to keep all importance weights well controlled. If there is a substantial mismatch between the children of a node and itself then choosing a parametric path between the two, adaptively specifying a tempering sequence along that path and using SMC techniques to approximate each distribution in turn may mitigate some difficulties (See Appendix~\ref{app:tempering}).

While we believe that the choice of the best $f_{t,u}, g_{t,u}$ will be model dependent, we found that, in some cases, some choices are preferable. As an example, take $f_t$ to be a Gaussian distribution with mean $\mu$ and covariance $\Sigma$. Then, a possible choice for $\gamma_{t,u}$ are Gaussian distributions with mean $\mu_u$ equal to the restriction of $\mu$ to the components corresponding to node $u$ and covariance $\Sigma_u$ obtained by subsetting $\Sigma$ and discarding all components not in $u$.
In this context, we found that setting $\Sigma_u = \Sigma(\mathcal{V}_u)^{-1}$, i.e. selecting the components of $\Sigma$ corresponding to node $u$ and then inverting this matrix, instead of $\Sigma_u = \Sigma^{-1}(\mathcal{V}_u)$ has lower computational cost and leads to more diffuse distributions and thus better behaved mixture weights~\eqref{eq:w_mix}.

\subsection{The Algorithm}

For each time $t$, having identified the space decomposition over $\mathbb{T}$ and the collection of distributions $(\widetilde{\gamma}_{t, u})_{ u\in\mathbb{T}}$, we can apply Algorithm~\ref{alg:dac} to the root $\mathfrak{R}$ of $\mathbb{T}$. However, since the integral in~\eqref{eq:gamma} is not analytically tractable, we replace $\widetilde{\gamma}_{t-1,\mathfrak{R}}$ with an approximation provided by the particle population at the root of the tree corresponding to $t-1$, $\{z_{t-1, \mathfrak{R}}^n\}_{n=1}^N$, i.e. its particle approximation obtained at the previous time step, as is normally done in MPFs, and define
\begin{align}
\label{eq:gamma_marginal}
    \gamma_{t,u}(z_{t,u}) &:= 
    g_{t,u}(z_{t,u}, (y_t(i))_{i\in\mathcal{V}_{u}}) \frac{1}{N}\sum_{n=1}^N f_{t,u}(z_{t-1, \mathfrak{R}}^n, z_{t,u}).
\end{align}

Given $\{z_{t-1, \mathfrak{R}}^n\}_{n=1}^N$, at each leaf node of the tree we sample one component of $x_t$ per node from $N^{-1}\sum_{n=1}^NK_{t,u}(z_{t-1, \mathfrak{R}}^n, \cdot)$, the importance weights are then given by
\begin{align}
\label{eq:w_leaf}
    w_{t, u}(z_{t,u}, x_{1:t-1,u})
    &=\frac{g_{t,u}(z_{t,u}, (y_t(i))_{i\in\mathcal{V}_u})\sum_{n=1}^Nf_{t,u}(z_{t-1, \mathfrak{R}}^n, z_{t,u})}{\sum_{n=1}^NK_{t,u}(z_{t-1, \mathfrak{R}}^n, z_{t,u})}.
\end{align}
As in the MPF, if we choose $K_{t,u}=f_{t,u}$, \eqref{eq:w_leaf} simplifies dramatically to $w_{t,y}(z_{t,u},x_{1:t-1,u}) = g_{t,u}(z_{t,u}, (y_t(i))_{i\in\mathcal{V}_u})$, considerably reducing the cost of evaluating the weights at the leaves.

For any non-leaf node $u$ we gather the particle populations $\{z_{t, \ell(u)}^n\}_{n=1}^N$ and $\{z_{t, r(u)}^n\}_{n=1}^N$ on its left and right child, respectively, and obtain an approximation of the product measure $ \gamma_{t, \mathcal{C}_u}:= \gamma_{t, \ell(u)}\times  \gamma_{t, r(u)}$ using the weighted product form estimator~\eqref{eq:pf}
\begin{align}
\label{eq:gamma_prod}
\gamma_{t, \mathcal{C}_u}^N = N^{-2}\sum_{n_1=1}^N\sum_{n_2=1}^N  w_{t, \ell(u)}(z_{t, \ell(u)}^{n_1})w_{t, r(u)}(z_{t, r(u)}^{n_2})\delta_{(z_{t, \ell(u)}^{n_1}, z_{t, r(u)}^{n_2})},
\end{align}
or one of the lower cost alternatives discussed in Section~\ref{sec:light}.

As in the DaC-SMC setting, we reweight the particle approximation of $\gamma_{t,\mathcal{C}_u}$ to target $\gamma_{t,u}$.
In this case, the mixture weights are given by
\begin{align}
\label{eq:w_mix}
m_{t, u}(z_{t, \mathcal{C}_u}) =& \frac{\gamma_{t, u}( z_{t,\mathcal{C}_u})}{\gamma_{t, \mathcal{C}_u}(z_{t, \mathcal{C}_u})}\\
=&\frac{g_{t,u}(z_{t, \mathcal{C}_u}, (y_t(i))_{i\in\mathcal{V}_u})}{g_{t,\ell(u)}(z_{t,\ell(u)}, (y_t(i))_{i\in\mathcal{V}_{\ell(u)}})g_{t,r(u)}(z_{t,r(u)}, (y_t(i))_{i\in\mathcal{V}_{r(u)}})} \times\notag \\
&\frac{N^{-1}\sum_{n=1}^N f_{t,u}(z_{t-1, \mathfrak{R}}^n, z_{t, \mathcal{C}_u})}{ N^{-1}\sum_{n=1}^Nf_{t,\ell(u)}(z_{t-1, \mathfrak{R}}^n, z_{t,\ell(u)}) N^{-1}\sum_{n=1}^N f_{t,r(u)}(z_{t-1, \mathfrak{R}}^n, z_{t,r(u)})}\notag,
\end{align}
where we defined $z_{t,\mathcal{C}_u}:=(z_{t,\ell(u)}, z_{t,r(u)})$ the vector obtained by merging the components on the left and on the right child of $u$.

For each pair in~\eqref{eq:gamma_prod} we obtain the incremental mixture weights $m_{t,u}^{(n_1, n_2)}:=m_{t,u}((z_{t,\ell(u)}^{n_1},z_{t,r(u)}^{n_2}))$ in~\eqref{eq:w_mix} and the updated weights
\begin{align*}
\tilde{w}_{t,u}^{(n_1, n_2)}=\tilde{w}_{t,u}((z_{t,\ell(u)}^{n_1},z_{t,r(u)}^{n_2})):= w_{t,\ell(u)}(z_{t,\ell(u)}^{n_1})w_{t,r(u)}(z_{t,r(u)}^{n_2})m_{t,u}((z_{t,\ell(u)}^{n_1},z_{t,r(u)}^{n_2})),
\end{align*}
for $n_1, n_2 = 1, \dots, N$.
To avoid unbounded growth in the number of particles, the weights $\tilde{w}_{t,u}$ are then used to resample a population of $N$ particles approximating $\gamma_{t, u}$, $\{\tilde{z}_{t,u}^n, w_{t,u}=1\}_{n=1}^N$ --- although one could allow the number of particles retained to grow as the simulation approaches the root of the tree to accommodate the growing dimension of the space.
Algorithm~\ref{alg:dac_filter} summarizes the procedure described above with the natural modification at $t=1$, where it is not necessary to ``marginalise'' over previous states and simple importance sampling can be used at the leaf nodes. Contrary to Algorithm~\ref{alg:dac}, we do not include an additional MCMC step in this statement of the algorithm, but one could easily be added as discussed in Section~\ref{sec:proposal}. In this case, given a $\pi_{t,u}$-invariant kernel $Q_{t,u}$, with $\pi_{t,u}\propto \gamma_{t,u}$, line $8$ should be replaced by: draw $ z_{t,u}^{n}\sim Q_{t,u}(\tilde z_{t,u}^{n},\cdot)$ for all $n\leq N$.

\begin{algorithm}[th]
\begin{algorithmic}[1]
\FOR{$u$ leaf node}
\STATE{\textit{Initialize:} draw $ z_{t,u}^{n}\sim N^{-1}\sum_{n=1}^NK_{t,u}(z_{t-1, \mathfrak{R}}^{n},\cdot)$ and compute $w_{t,u}^{n}$ as in~\eqref{eq:w_leaf} for all $n\leq N$.}
\ENDFOR
\FOR{$u$ non-leaf node}
\STATE{\textit{Recurse:} set $(\{z_{t,v}^{n}, w_{t,v}^{n}\}_{n=1}^N):=\text{dac\_smc}(t, v)$ for $v$ in $\{\ell(u), r(u)\}$ and obtain $\gamma_{\mathcal{C}_u}^N$ in~\eqref{eq:gamma_prod}.}
\STATE{\textit{Merge:} compute the mixture weights $m_{t,u}^{(n_1, n_2)}$ in~\eqref{eq:w_mix} and $\tilde{w}_{t,u}^{(n_1, n_2)}$ for all $n_1, n_2 \leq N$.}
\STATE{\textit{Resample:} draw $\{ \tilde{z}_{t,u}^{n}\}_{n=1}^N$ using weights $\tilde{w}_{t,u}^{(n_1, n_2)}$ and set $w_{u}^{n}=1$ for all $n\leq N$.}
\STATE{\textit{Update:} set $z_{t,u}^{n}=\tilde{z}_{t,u}^{n}$ for all $n\leq N$.}
\ENDFOR
\STATE{\textit{Output} $(\{z_{t, \mathfrak{R}}^{n}\}_{n=1}^N)$.}
 \end{algorithmic}
 \caption{dac\_smc$(t)$ for $t\geq 1$. Given $(\{z_{t-1, \mathfrak{R}}^{n}\}_{n=1}^N):=\text{dac\_smc}(t-1)$.}\label{alg:dac_filter}
\end{algorithm}

The mixture resampling strategy described above requires evaluating the mixture weights~\eqref{eq:w_mix} for each of the $N^2$ pairs in~\eqref{eq:gamma_prod}. 
The $O(N^2)$ cost of this operation is often prohibitive for large $N$; we describe smaller cost alternatives in Section~\ref{sec:light} and demonstrate that these approaches have good performance in Section~\ref{sec:expe}.

\subsection{Choice of Proposals}
\label{sec:proposal}

Algorithm~\ref{alg:dac_filter} describes a general strategy to perform filtering using DaC-SMC; as in the case of standard SMC the performances of the algorithm are heavily influenced by the choice of the proposals at the leaf nodes. We discuss here a simple strategy to select $K_{t,u}$.

We assume that we can sample from $f_{t,u}$ in~\eqref{eq:gamma} and set $ K_{t,u}(z_{t-1, \mathfrak{R}}^n,\cdot)=f_{t,u}(z_{t-1, \mathfrak{R}}^n,\cdot)$ so that the importance weights~\eqref{eq:w_leaf} reduce to $w_{t, u} = g_{t,u}$. This choice corresponds to the proposal used in the bootstrap PF of \citet{gordon1993novel}, (locally) optimal proposals also exist (see e.g., \citet[Chapter 10]{chopin2020}) and are expected to lead to better performances but incur a higher computational cost.

While picking $ K_{t,u}(z_{t-1, \mathfrak{R}}^n,\cdot)=f_{t,u}(z_{t-1, \mathfrak{R}}^n,\cdot)$ causes standard marginal PFs to reduce to the bootstrap PF (as described in \citet[Section 3.3]{klaas2012toward}), this is not true for our marginal DaC-SMC, since the integral w.r.t. $z_{t-1, \mathfrak{R}}$ still appears in the mixture weights~\eqref{eq:w_mix}.

If needed, to avoid particle impoverishment, one might consider applying a Markov kernel $Q_{t,u}$ which leaves $\pi_{t,u}\propto\gamma_{t,u}$ invariant after the resampling step in line 7 of Algorithm~\ref{alg:dac_filter}.
These $\pi_{t,u}$-invariant kernels can be selected exploiting the vast literature on sequential MCMC methods (e.g., \citet{gilks2001following, septier2009mcmc, carmi2012gaussian, septier2015langevin, pal2018sequential, han2021application}) to employ proposals whose cost is not $O(N)$ as it would be for some na{\"\i}ve choices.

\subsection{Adaptive Lightweight Mixture Resampling}
\label{sec:light}
The mixture resampling in line 6--7 of Algorithm~\ref{alg:dac_filter} becomes computationally impractical for large $N$, since it requires evaluating the mixture weights for $N^2$ particles \citep{lindsten2017divide, Kuntz2021, Corneflos2021}.
Several strategies have been proposed to alleviate this cost by only constructing a subset of the $N^2$ combinations in~\eqref{eq:gamma_prod}, e.g. the multiple matching strategy of \cite{lin2005independent} which gives rise to the lightweight mixture resampling of \citet{lindsten2017divide} in this context, strategies borrowed from the literature on incomplete U-statistics \citep{Kuntz2021} and lazy resampling schemes \citep{Corneflos2021}.

We consider the lightweight version of mixture resampling proposed in \citet{lindsten2017divide} which only considers a subset $\theta N$ with $\theta\ll N$ of the $N^2$ possible pairs. However, instead of setting $\theta$ to some pre-specified value (e.g. $\theta= \lceil\sqrt{N}\rceil$),
we propose a simple strategy to select $\theta$ adaptively based on the effective sample size ($\ess$; \cite{kong1994sequential}),
\begin{align}
\label{eq:ess}
\ess:= \left(\sum_{n} \tilde{w}_{t,u}^{n}\right)^2 /\sum_{n} (\tilde{w}_{t,u}^{n})^2,
\end{align}
where the sum is over all pairs $n=(n_1, n_2)$ obtained from~\eqref{eq:gamma_prod},
which is similar in spirit to the adaptive tempering strategies commonly encountered in the SMC literature (see, e.g., \cite{Jasra2010,johansen2015blocks}) and aiming to do just enough computation to obtain a good $N$-sample approximation.

The merge step in lines 6--7 of Algorithm~\ref{alg:dac_filter} is replaced by Algorithm~\ref{alg:adaptive_light}: after building all the $N$ pairs obtained by concatenating the particles associated with each of the two children, further permutations are added until the $\ess$
achieves a pre-specified value $\ess^\star$ (e.g.  $\ess^\star=N$). To avoid $\theta$ getting too large, we stop adding permutations when $\theta= \lceil\sqrt{N}\rceil$ thereby allowing us to bound the worst-case computational cost by $O(N^{3/2})$. 
We empirically compare several mixture resampling approaches in Appendix~\ref{app:resampling}. 

\begin{algorithm}[th]
\begin{algorithmic}[1]
\STATE{\textit{Correct:} compute the mixture weights $m_{t,u}(x_{\ell(u)}^n,x_{r(u)}^n)$ as in~\eqref{eq:w_mix} and $\tilde{w}_{t,u}(x_{\ell(u)}^n,x_{r(u)}^n)$ for all $n\leq N$ and the $\ess$~\eqref{eq:ess}.}
\STATE{\textit{Set:} $\theta \leftarrow 1$ and $\tilde{x}_u^n = (x_{\ell(u)}^n,x_{r(u)}^n)$ for $n \leq N$.}
\WHILE{$\ess((\tilde{x}^n)_{n=1}^{\theta N}) <\ess^\star$ and $\theta<\left\lceil\sqrt{N}\right\rceil$}
\STATE{\textit{Set:} $\theta \leftarrow \theta+1$.}
\STATE{\textit{Permute:} draw one permutation of $N$, $\pi(N)$, set $\tilde{x}_u^{N(\theta-1)+n} = (x_{\ell(u)}^n,x_{r(u)}^{\pi(n)})$, compute the mixture weights $m_{t,u}(\tilde{x}_u^{N(\theta-1)+n})$ in~\eqref{eq:w_mix} and the updated weights $\tilde{w}_{t,u}(\tilde{x}_u^{N(\theta-1)+n})$ for $n \leq N$ and update the $\ess$.}
\ENDWHILE
\STATE{\textit{Resample:} draw $\{ z_{t,u}^{n}\}_{n=1}^N$ from $\tilde{x}_u^{1:N\theta}$ with weights $\tilde{w}_{t,u}(\tilde{x}_u^{1:N\theta})$ and set $w_{t,u}^{n}=1$ for all $n\leq N$.}
\end{algorithmic}
\caption{Adaptive lightweight mixture resampling.}\label{alg:adaptive_light}
\end{algorithm}

Algorithm~\ref{alg:adaptive_light} can be implemented in a space-efficient manner by storing the permutations $\pi(N)$ corresponding to each value of $\theta$ rather than building the $\theta N$ pairs $\tilde{x}_u^{1:N\theta}$.

\begin{figure}
\centering
\begin{tikzpicture}[every node/.append style={font=\normalsize}]
\node (img1) {\includegraphics[width = 0.6\textwidth]{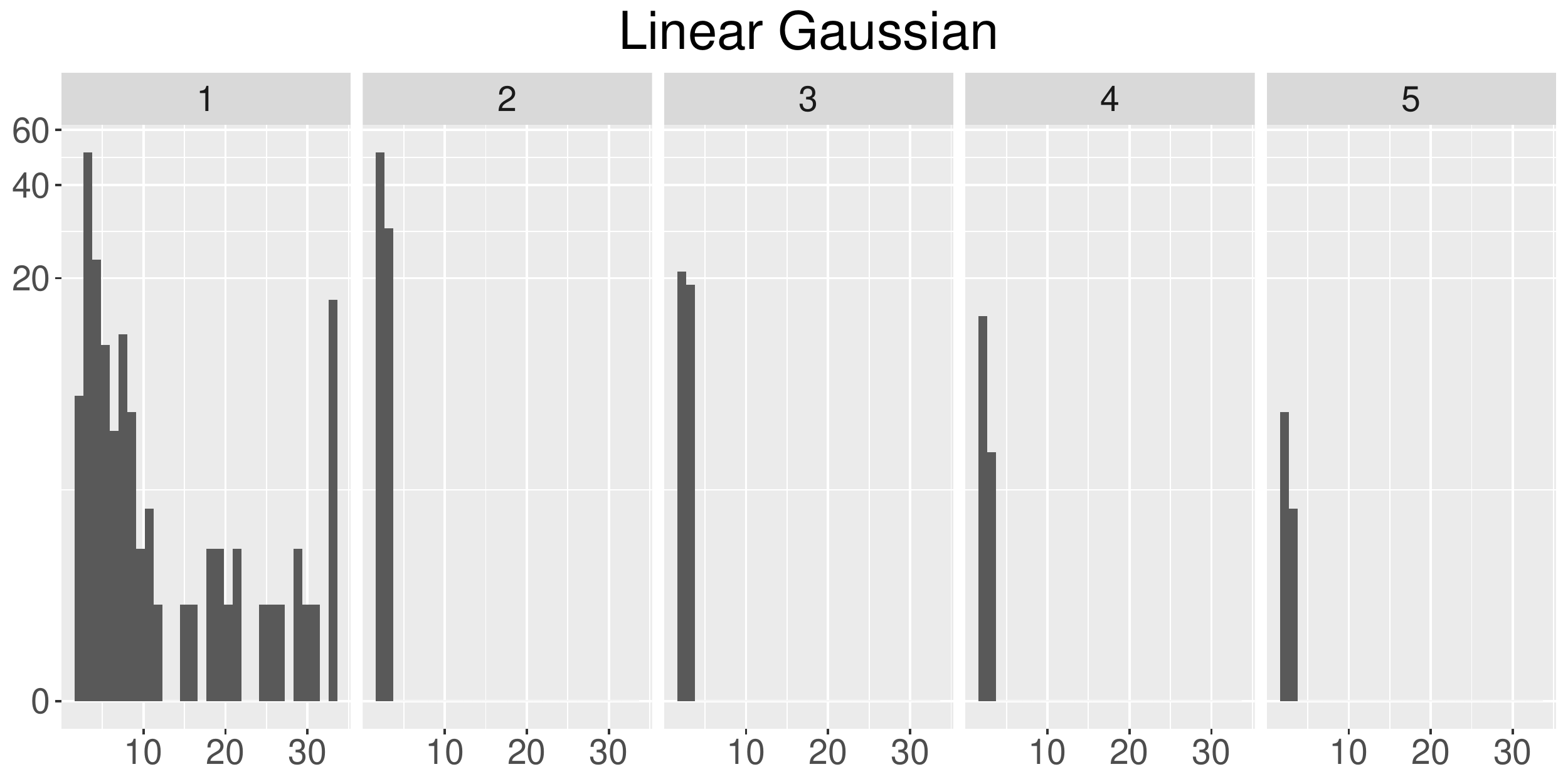}};
\end{tikzpicture}
\caption{Distribution of the number of permutations $\theta$ selected by Algorithm~\ref{alg:adaptive_light} for the simple linear Gaussian model in Section~\ref{sec:lgssm} with $d=32$, $N=10^3$ and 10 time steps; the 5 panels correspond to the 5 levels of the tree from the level above the leaves (level 1) to the root (level 5).}
\label{fig:m_adaptive}
\end{figure}

We report in Figure~\ref{fig:m_adaptive} the distribution of the number of permutation $\theta$ selected by Algorithm~\ref{alg:adaptive_light} for the linear Gaussian model in Section~\ref{sec:expe}.
We observe that the highest values of $\theta$ are chosen at the level above the leaves (panel $1$), indeed, at the leaf level the observation $y_t$ is incorporated, causing a larger adjustment to the distribution at the first mixture resampling step than that needed as we move up the tree.
The spike at $\theta=\lceil\sqrt{N}\rceil$ at level 1 shows that sometimes the target $\ess$ is not reached, which suggests that the product of the proposal distributions over the children might be a poor proposal for $\gamma_{t,u}$ --- this is in part likely to be due to the use of a ``bootstrap proposal'' and could be mitigated in the same way as standard particle filters by seeking to design (marginal) proposal distributions which incorporate the influence of observations.
In addition, this phenomenon can be mitigated with the use of (adaptive) tempering, as shown in, e.g., \citet{Jasra2010,johansen2015blocks, wang2020annealed, zhou2016toward} for standard SMC and \citet[Section 4.2]{lindsten2017divide} for DaC-SMC. However, na{\"\i}ve implementation of this approach will bear a substantial computational cost in the marginal context; designing MCMC kernels that are efficient in this context has been explored in the sequential MCMC context --- see, e.g., \citet[Section III-B]{septier2015langevin} --- and in marginal STPFs \citep{xu2019particle}. We give details of a tempering strategy for Algorithm~\ref{alg:adaptive_light} in Appendix~\ref{app:tempering}. 

\subsection{Computational Cost}
\label{sec:cost}
At a given node, $u$, the runtime cost of Algorithm~\ref{alg:dac_filter} with lightweight mixture resampling given the particle approximations for its children, is $O(h(N, d)\theta N)$ where $h(N, d)$ denotes the cost of obtaining the mixture weights~\eqref{eq:w_mix}.

In general, the dependence of $h(N, d)$ on the number of particles $N$ is $O(N)$ as for MPFs. Similarly to MPFs, one could try to reduce the cost of computing~\eqref{eq:w_mix} making use of techniques from $N$-body learning (e.g. \cite{gray2000n, lang2005empirical}) as shown in \cite{klaas2012toward}. Alternatively, one could consider efficient implementations using GPUs \citep{charlier2021kernel} as shown in \cite[Section 4]{clarte2019collective} for sums of the form of those in~\eqref{eq:w_mix}.

For some models, it might be possible to pick the auxiliary functions $f_{t,u}$ so that the dependence on the past (i.e. $z_{t-1, \mathfrak{R}}$ in~\eqref{eq:gamma_marginal}) vanishes and obtain an $O(1)$ cost w.r.t. $N$. However, we expect that this type of decompositions will require larger corrections at the root, where $f_{t, \mathfrak{R}}=f_t$, which might offset the cost savings.

In the adaptive case, worst case costs are given by $O(h(N, d)\theta N)$ with $\theta$ replaced by the upper bound imposed upon the number of permutations considered. For instance, for the examples in Section~\ref{sec:expe} we have an $h(N, d)=O(N)$ cost to obtain the mixture weights and we set the upper bound for $\theta$ to be $N^{1/2}$ leading to a cost of order $O(N^{5/2})$ w.r.t. the number of particles.

Denoting $C_u(d,\theta_u, N)$ the cost of running Algorithm~\ref{alg:dac_filter} at node $u$, we can then bound the total cost of serial implementations Algorithm~\ref{alg:dac_filter} applied at the root node $\mathfrak{R}$ by $O(dt\sup_u C_u)$, where the supremum is taken over all nodes in $\mathbb{T}$; a lower running cost of $O( t \sup_u C_u\log_2 d)$, can be achieved parallelizing the computations over each level of the tree \citep[Section 5.3]{lindsten2017divide}. 

In the adaptive case, this upper bound is far from being tight, since, as shown in the histogram in Figure~\ref{fig:m_adaptive}, $\theta$ tends to be high when the observation is incorporated (level 1) but stabilizes as we move up the tree.
Additionally, for large $N$ one expects the number of permutations required to obtain a good $N$-particle approximation to converge to some fixed integer and hence the cost for sufficiently large $N$ will with high probability be of smaller order than these bounds. Furthermore the constants multiplying the $N^{5/2}$ contribution arising from the level above the leaves are sufficiently small that this is not the dominant cost in our experiments --- and is likely to be typical in high-dimensional settings in which it is rarely feasible to employ very large numbers of particles and the objective is to obtain a good approximation at acceptable time \emph{and space} costs.

\section{Experiments}
\label{sec:expe}
We compare the results obtained with DaC with those of NSMC and STPF; we do not include simpler strategies because both standard PF and BPF have been shown to have worse performances than NSMC and STPF for the model considered here \citep{naesseth2015nested, naesseth2019high, beskos2017stable}, and the marginal version of STPF because of the higher cost for large $d$.
 
The functions $g_{t,u}, f_{t,u}$ in~\eqref{eq:gamma} are obtained from $g_t, f_t$, respectively, by discarding all the terms in those functions involving components $i\not\in\mathcal{V}_{u}$, further details are given in Appendix~\ref{app:details}.
For Algorithm~\ref{alg:dac_filter} we use the proposals discussed in Section~\ref{sec:proposal} and the lightweight mixture resampling strategies described in Section~\ref{sec:light}.
All resampling steps are performed using stratified resampling \citep{kitagawa1996monte}.

First, we consider a simple linear Gaussian SSM, and compare the results obtained by the three algorithms with the exact filtering distribution given by the Kalman filter. Then, we consider a spatial model with simple latent dynamics but non-trivial spatial correlations structure between observations, moving away from the assumption of i.i.d. observations which is convenient from a computational perspective, but rarely satisfied in practice \citep{chib2009multivariate}.
  
All the experiments have been executed in serial using a single core of a Intel(R) Xeon(R) CPU E5-2440 0 @ 2.40GHz using R 4.1.0.

\subsection{Simple Linear Gaussian Model}
\label{sec:lgssm}
We start by considering a simple linear Gaussian SSM taken from \cite{naesseth2015nested}, for which the filtering distributions can be computed exactly with the Kalman filter. The model is given by
$f_t(x_{t-1}, x_t) = \N(x_t; Ax_{t-1}, \Sigma)$, and $g_t(x_t, y_t) = \N(y_t; x_t,\sigma^2_y \Id_d)$,
with $A\in\real^{d\times d}$, $\sigma_y^2>0$, $\Sigma \in \real^{d\times d}$ a tridiagonal covariance matrix and $\Id_d$ the $d$-dimensional identity matrix (see Appendix~\ref{app:detailsgaussian} for full details and computation of the mixture weights~\eqref{eq:w_mix}).

We compare DaC with both non-adaptive and adaptive lightweight mixture resampling with 2-level NSMC with fully adapted outer level and STPF on data simulated from the model for $d=2^5, 2^8, 2^{11}$ for $t=100$ time steps. We use different number of particles $N=100, 500, 1000$ for Algorithm~\ref{alg:dac_filter} and the outer level of NSMC and STPF, while the number of particles for the inner level of NSMC and the number of particles for each island of STPF is fixed to $M=100$ as suggested in \cite{naesseth2015nested, beskos2017stable}.

To evaluate the results, we consider two global measures of accuracy for each of the $d$ marginals, the Wasserstein-1 distance (see, e.g., \cite{vallender1974}), and the Kolmogorov-Smirnov distance
\begin{align*}
W_{1, i} := \int \left\lvert F_{t, i}(x) - \widehat{F}_{t, i}(x)\right\rvert\rmd x,\qquad\qquad  \ks_i :=\max_x \left\lvert F_{t, i}(x) - \widehat{F}_{t, i}(x)\right\rvert, 
\end{align*}
where $ F_{t, i}$ denotes the 1-dimensional cumulative distribution function of marginal $i$ at time $t$ and $\widehat{F}_{t, i}$ its particle approximation. Further comparisons, which demonstrate that the mean squared error ($\mse$) of the filtering mean behaves similarly, are collected in Appendix~\ref{app:expegaussian}.

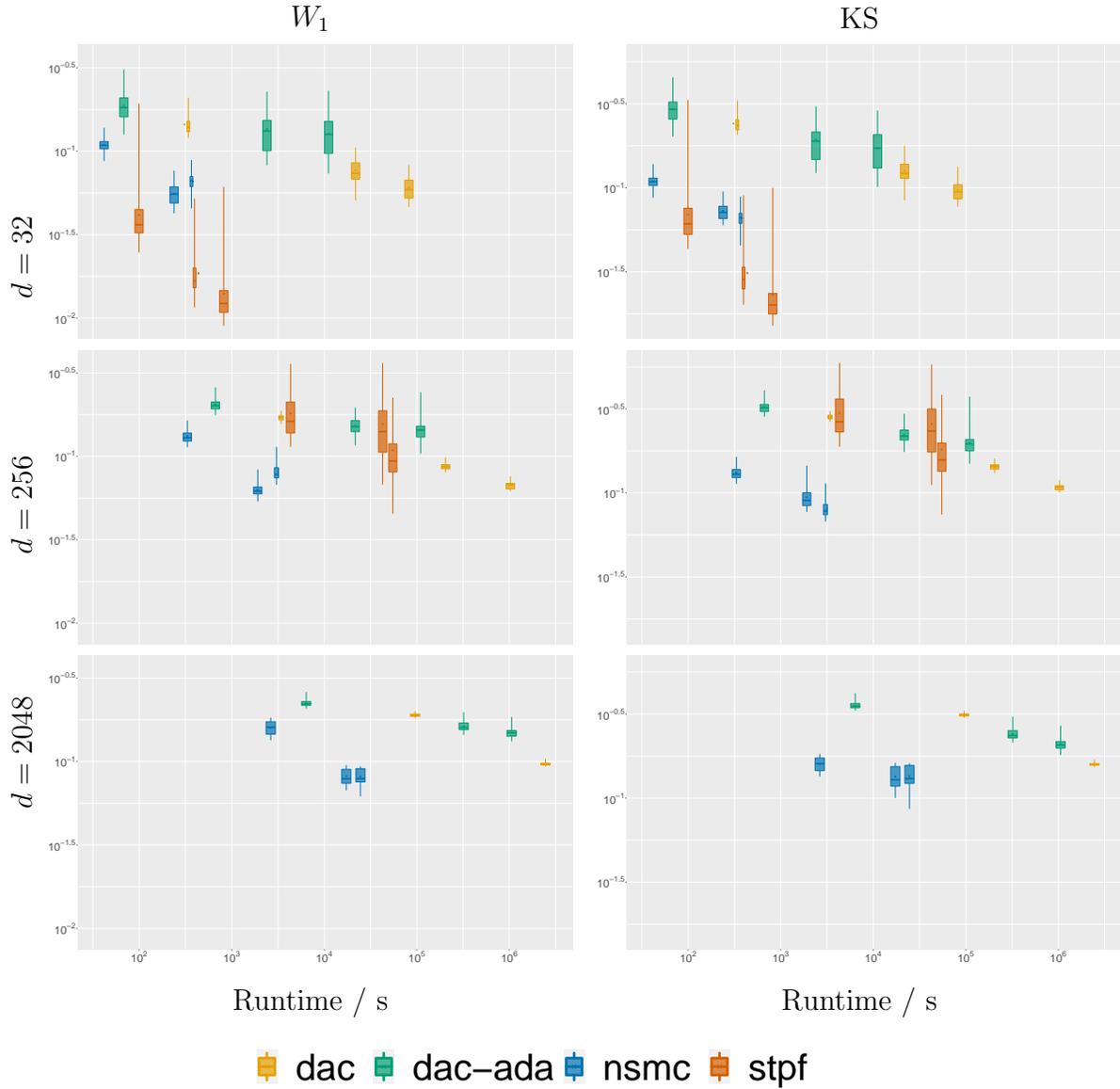
\begin{figure}
\centering
\begin{tikzpicture}[every node/.append style={font=\normalsize}]
\node (img1) {\includegraphics[width = 0.45\textwidth]{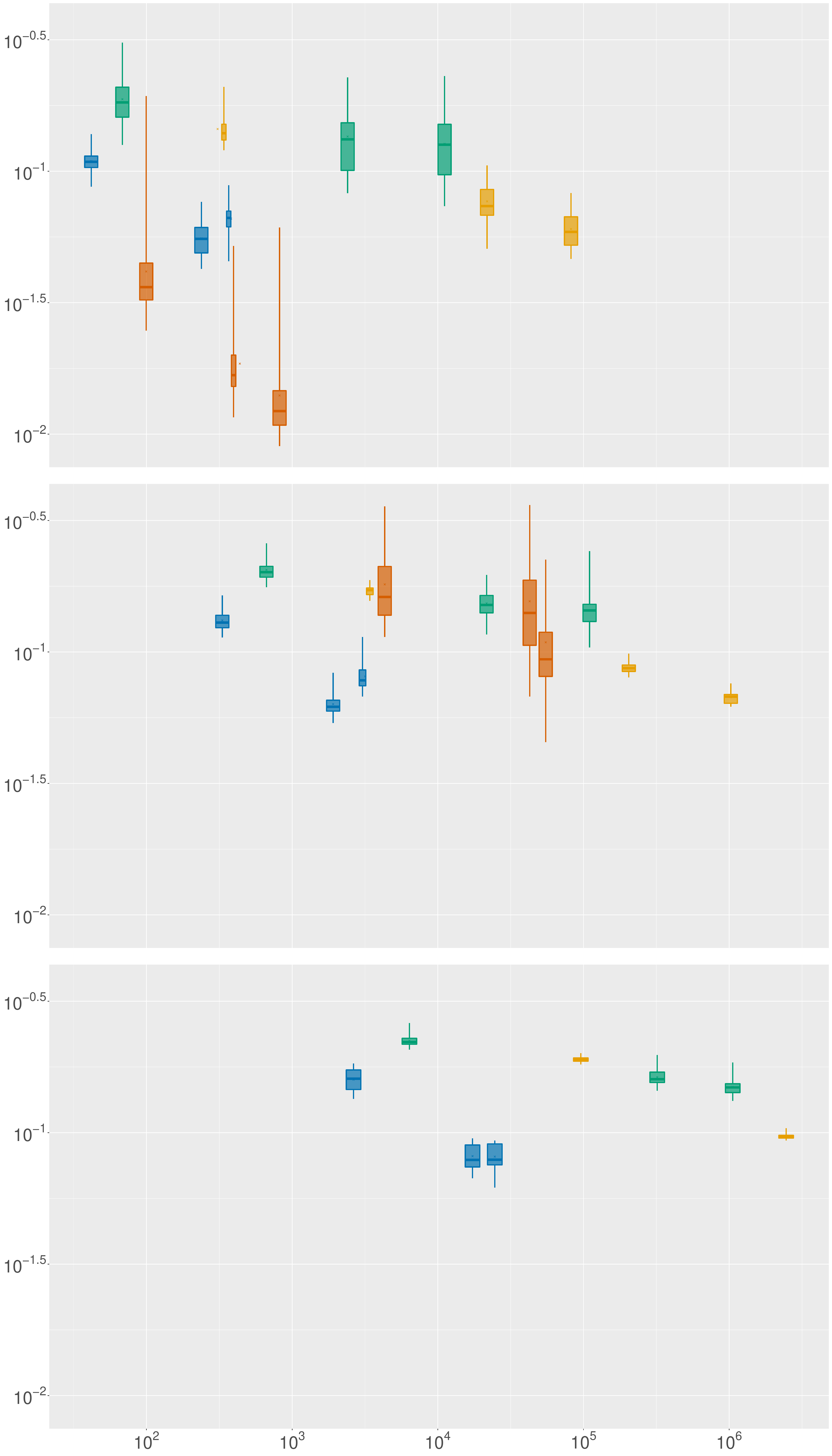}};
\node[above=of img1, node distance = 0, anchor = center, yshift = -0.8cm] {$W_1$};
\node[left=of img1, node distance = 0, rotate = 90, anchor = center, yshift = -0.8cm, xshift = 3.5cm] {$d=32$};
\node[left=of img1, node distance = 0, rotate = 90, anchor = center, yshift = -0.8cm] {$d=256$};
\node[left=of img1, node distance = 0, rotate = 90, anchor = center, yshift = -0.8cm, xshift = -3.5cm] {$d=2048$};
\node[right=of img1, xshift = -1cm] (img2) {\includegraphics[width = 0.45\textwidth]{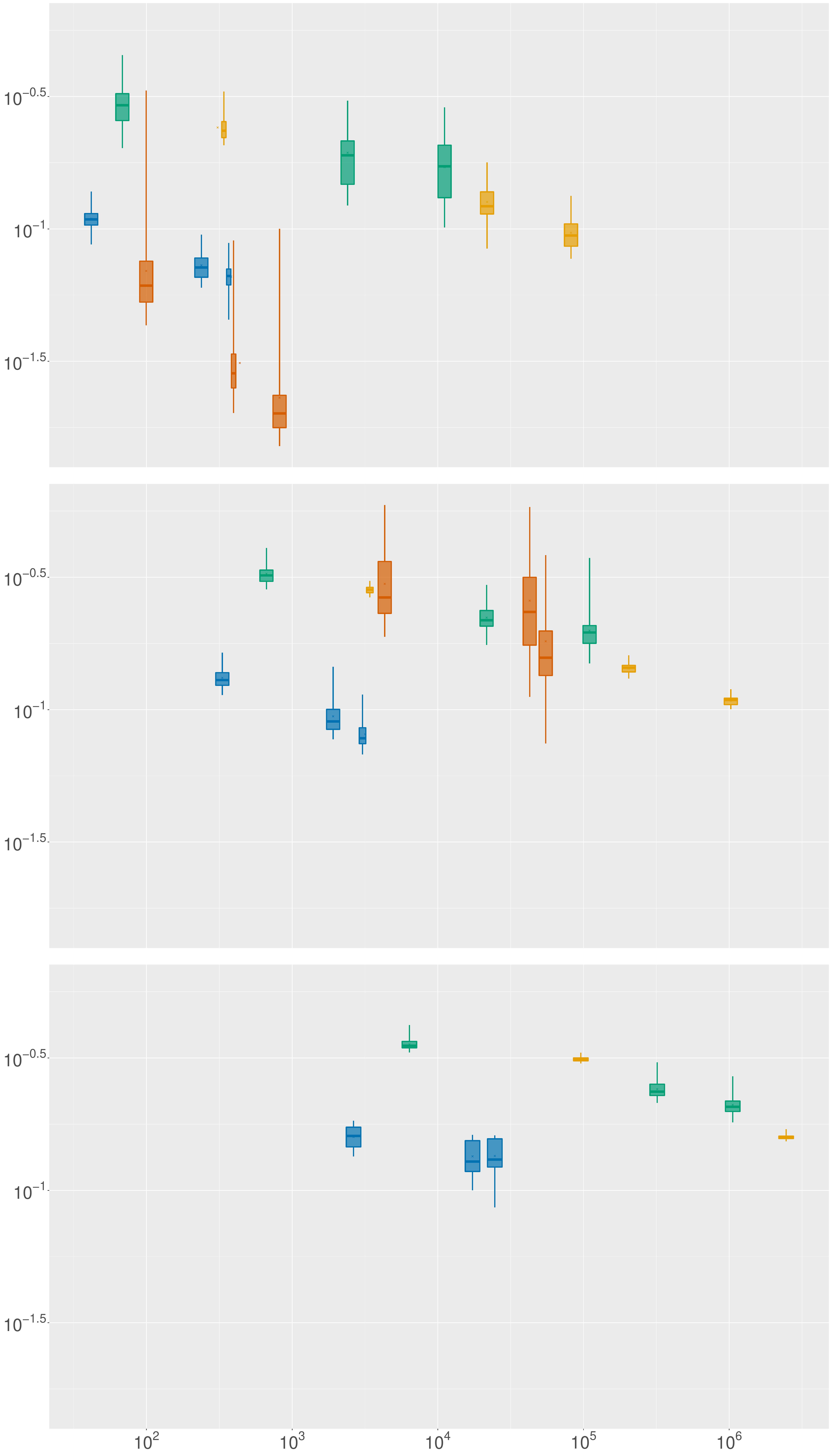}};
\node[above=of img2, node distance = 0, anchor = center, yshift = -0.8cm] {$\ks$};
\node[below=of img1, node distance = 0, yshift = 1cm] {Runtime / s};
\node[below=of img2, node distance = 0, yshift = 1cm] {Runtime / s};
\node[below=of img1, node distance = 0, xshift = 3cm, yshift = 0.5cm]{\includegraphics[width = 0.6\textwidth]{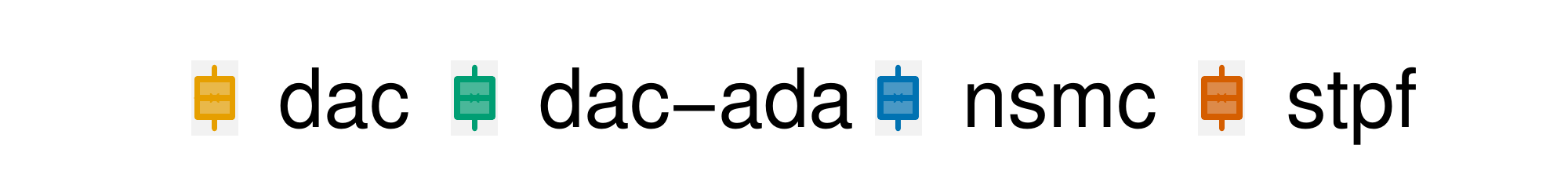}};
\end{tikzpicture}
\caption{Comparison of DaC, NSMC and STPF for $d=2^5, 2^8, 2^{11}$. Distribution of the average (over dimension) $W_1$ and $\ks$ distance at the last time step $t=100$ for 50 runs; the boxes, form left to right, correspond to increasing number of particles ($N=100, 500, 1000$). 
Due to their excessive cost, we do not include the results for STPF with $d=2048$ and those of the non-adaptive version of DaC $d=2048, N=1000$.}
\label{fig:lgssm}
\end{figure}

For lower-dimensional problems (e.g. $d=32$) STPF achieves the best results both in terms of Wasserstein-1 distance and in terms of $\ks$ distance (Figure~\ref{fig:lgssm}) and the relative $\mse$ of the reconstructions is considerably smaller (almost one order of magnitude smaller; see Appendix~\ref{app:expegaussian}). The results provided by STPF deteriorate quickly as $d$ grows, for $d=256$ the estimates of $W_1$ and $\ks$ distance are significantly worse than those provided by NSMC or DaC without adaptation.

The cost of STPF grows quadratically with $d$, and becomes unmanageable for large $d$, it is therefore not included in Figure~\ref{fig:lgssm} bottom panels.
STPF has the higher cost also for lower dimension (Figure~\ref{fig:lgssm} top panel), but in this case STPF provides the best results.
STPF has higher variability than the other methods, and even when the average results are better than DaC and NSMC (e.g. $d=32$), $W_1$ and $\ks$ can take considerably high values.

The results in terms of $\ks$ are generally more variable, this is likely due to the fact that $\ks$ is a measure of the worst case mismatch between $F_{t, i}(x)$ and $\widehat{F}_{t, i}(x)$, while for $W_1$ the mismatch is averaged over locations.
For large $d$, DaC has the smallest variability among the three algorithms.

DaC with fixed-cost lightweight mixture resampling generally gives better results than the adaptive lightweight mixture resampling, however, the cost of the latter is considerably smaller, making the adaptive version still manageable for large $N$ whereas the fixed-cost lightweight mixture resampling becomes too costly for large $N$ and large $d$. As discussed in Section~\ref{sec:cost}, the computational cost of both versions of DaC could be reduced using GPUs. 
In particular, both $W_1$ and $\ks$ decay more quickly with $N$ for DaC without adaptation than for DaC with adaptive lightweight mixture resampling.
The decay with $N$ is less evident for NSMC.
\subsection{Spatial Model}
\label{sec:ng}

We consider a model on a 2D-lattice in which the latent dynamics are simple but the observation structure is challenging.
The components of $X_t$ are indexed by the vertices $v\in V$ of a lattice, where $V=\{1, \dots, d\}^2$, and follow a simple linear evolution $X_t(v) = X_{t-1}(v)+U_t(v)$, where $U_t(v)\overset{\textrm{i.i.d}}{\sim}\N(0, \sigma_x^2)$.
The observations model is $
Y_t=X_t+V_t$, where we take $V_t$ to be jointly $t$-distributed with $\nu=10$ degrees of freedom, mean zero and precision structure encapsulating a spatial component.
Let $D$ denote the graph distance, then the entry in row $v$ and column $j$ of the precision matrix $\Sigma^{-1}$ is given by $(\Sigma^{-1})_{vj}=\tau^{D(j, v)}$ if $D(j, v)\leq r_y$ and 0 otherwise.
We obtain data from the model above with $\sigma^2_x=1$, $\tau = -0.25$, $r_y=1$ and $t=10$.
The observation density does not factorize, and therefore NSMC and STPF cannot be applied (at least without approximating $g$ with e.g. a Gaussian or discarding the covariance information). To validate the correctness of the algorithm, we compare the results obtained by DaC with those of the standard bootstrap PF in Appendix~\ref{app:expe} on a small lattice and found the agreement to be excellent.

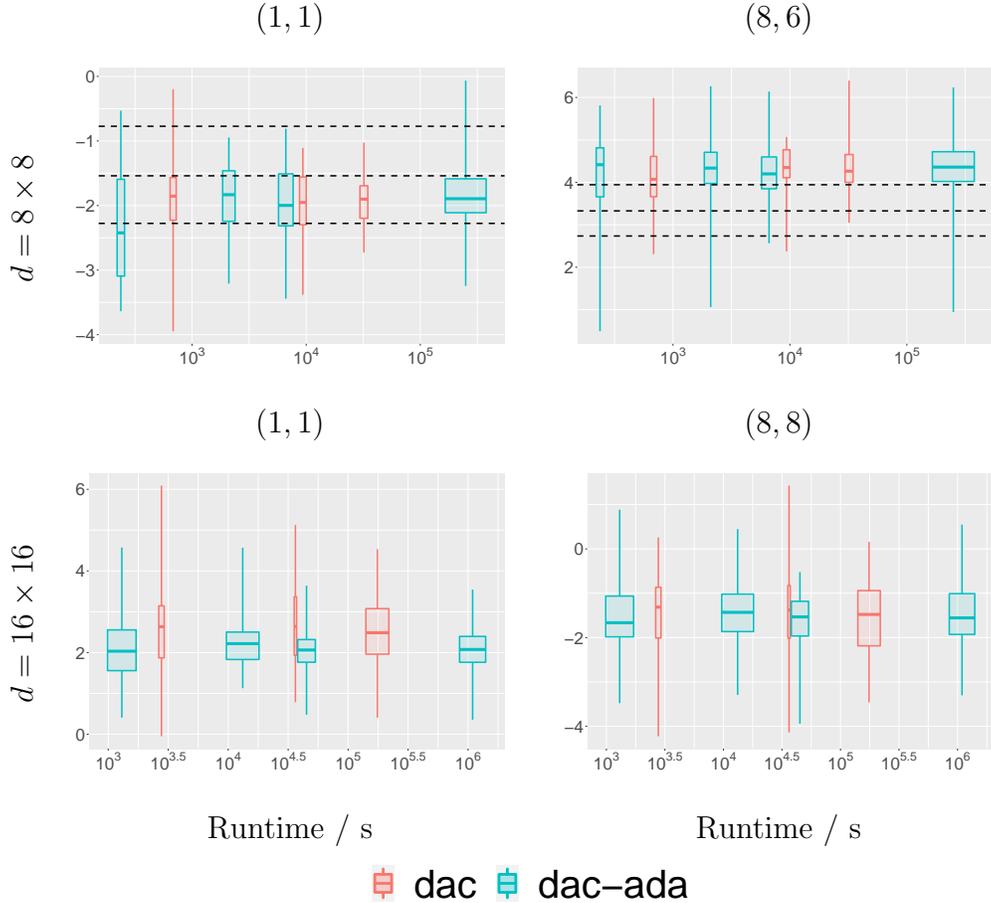
\begin{figure}
\centering
\begin{tikzpicture}[every node/.append style={font=\normalsize}]
\node (img1) {\includegraphics[width = 0.35\textwidth]{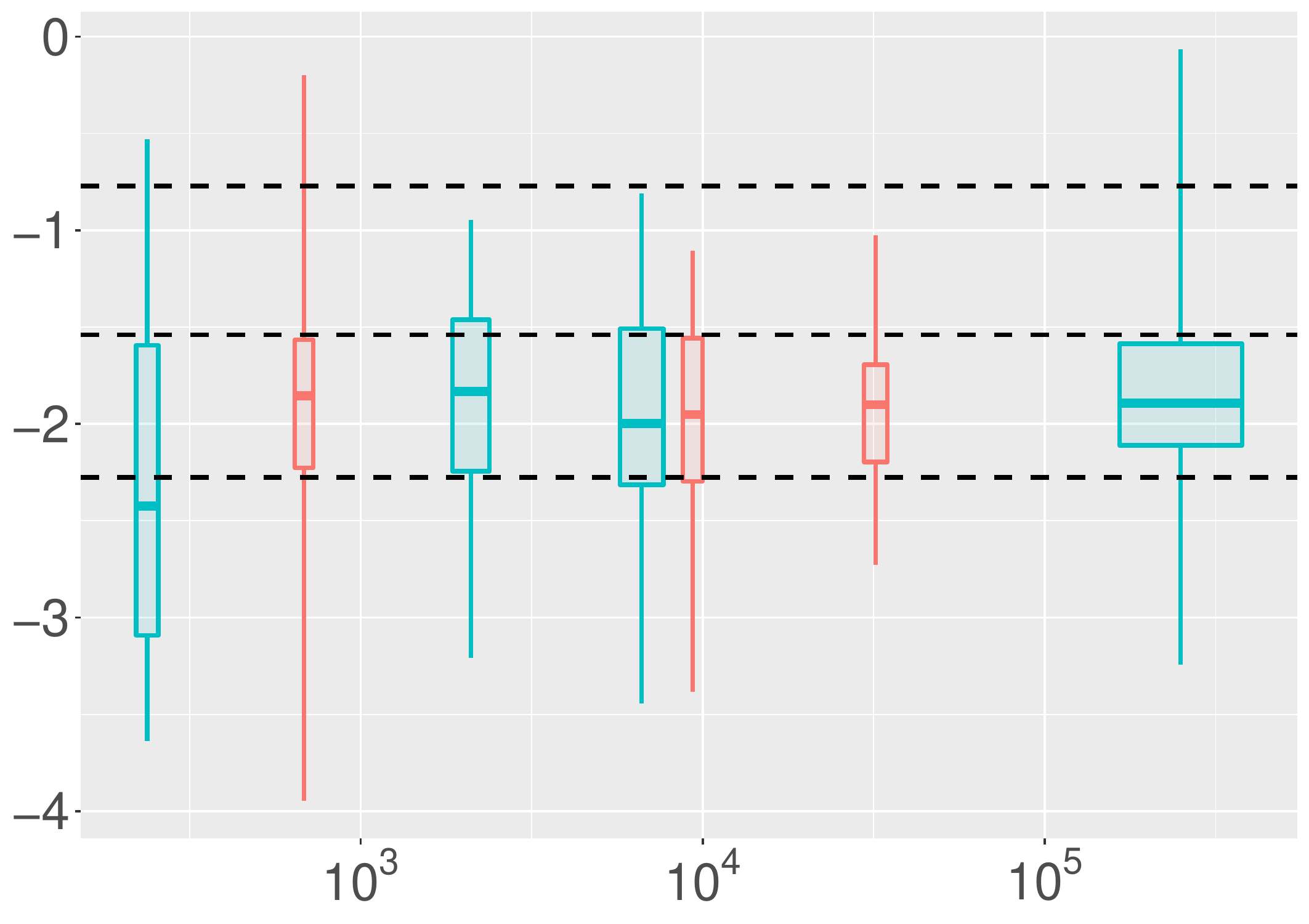}};
\node[right=of img1, xshift = -0.6cm] (img2)  {\includegraphics[width = 0.35\textwidth]{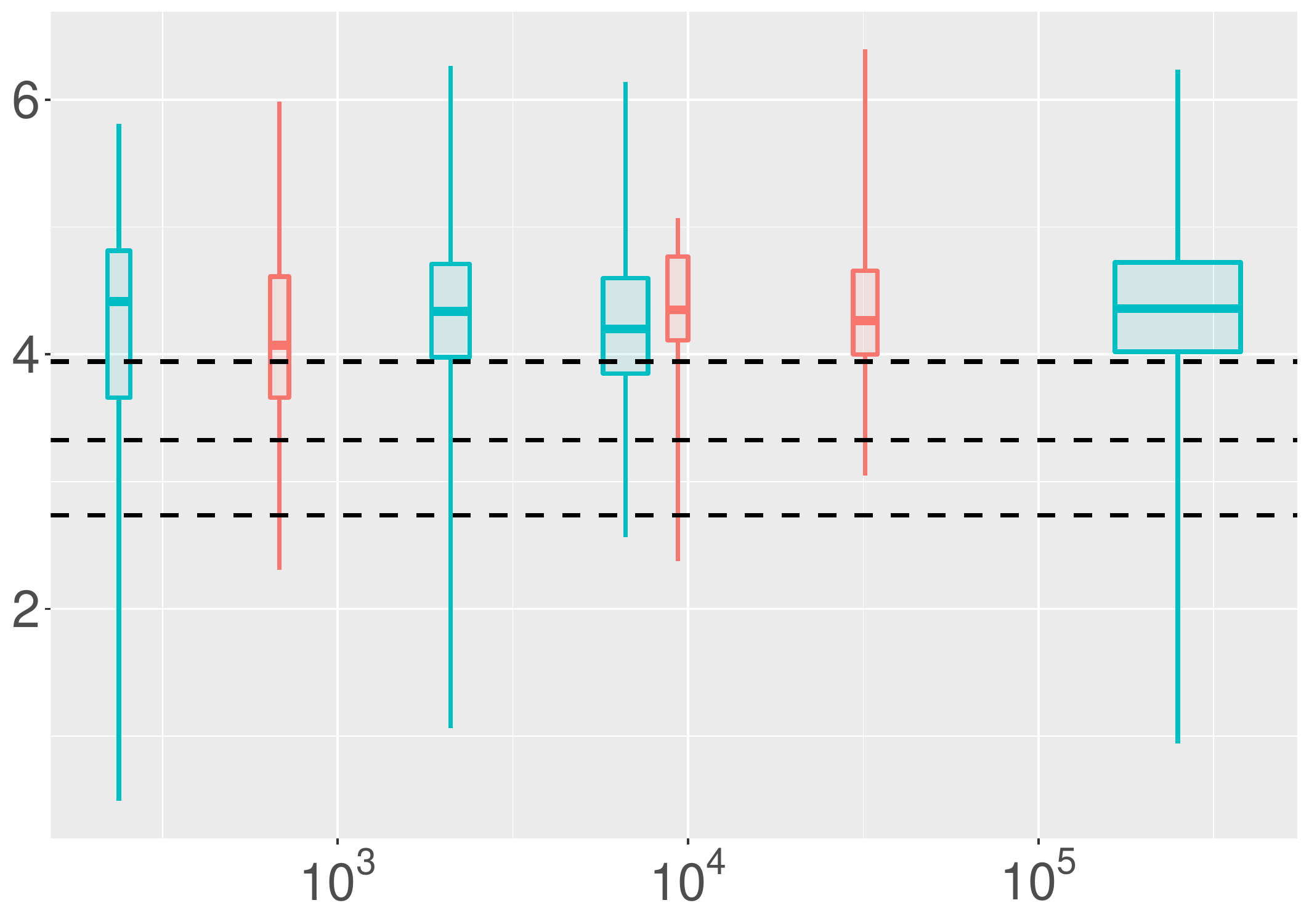}};
\node[above=of img1, yshift = -0.9cm] {$(1,1)$};
\node[above=of img2, yshift = -0.9cm] {$(8, 6)$};
\node[left=of img1, node distance = 0, rotate = 90, anchor = center, yshift = -0.5cm] {$d=8\times8$};
\node[below=of img1] (img3)  {\includegraphics[width = 0.35\textwidth]{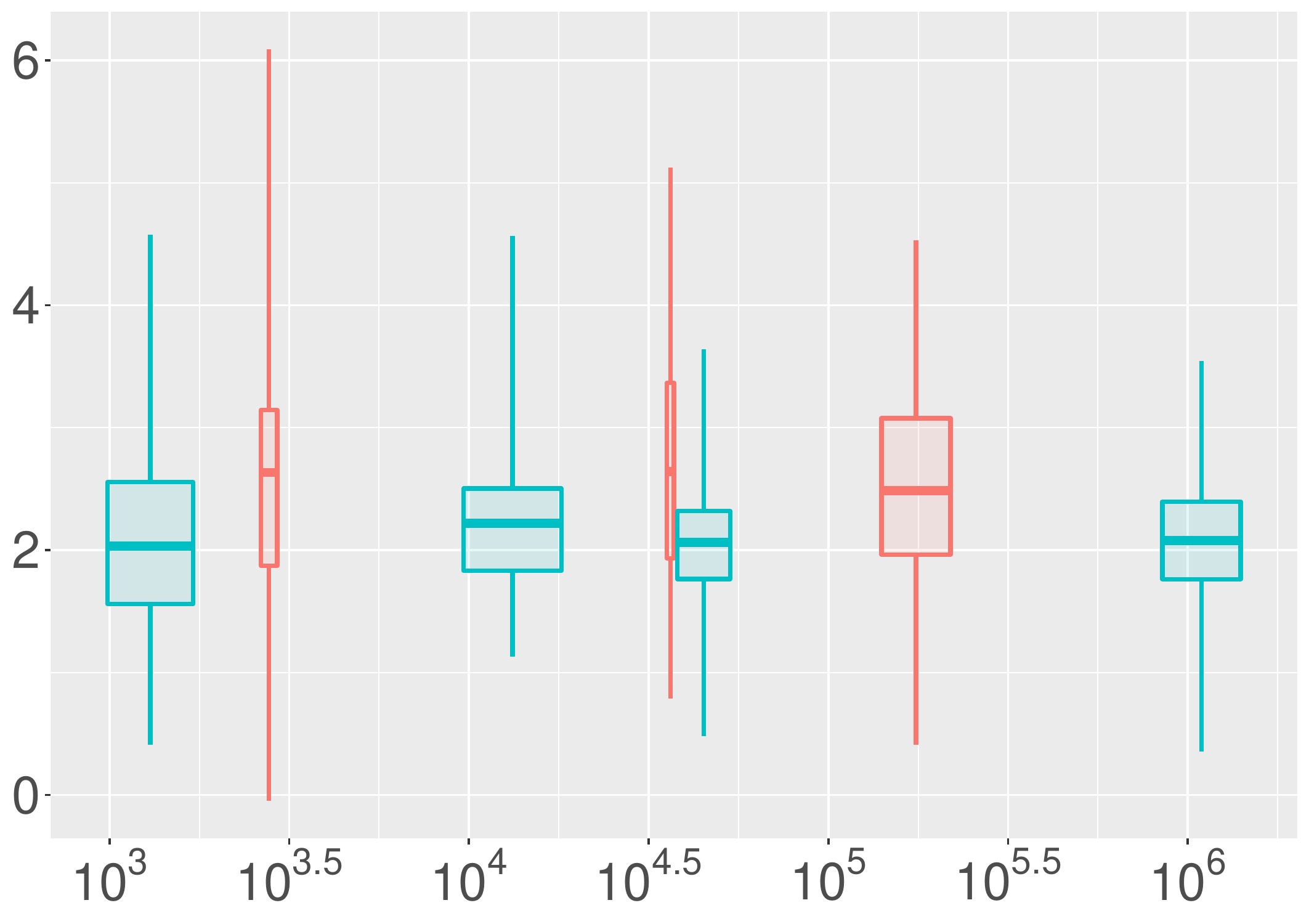}};
\node[right=of img3, xshift = -0.6cm] (img4)  {\includegraphics[width = 0.35\textwidth]{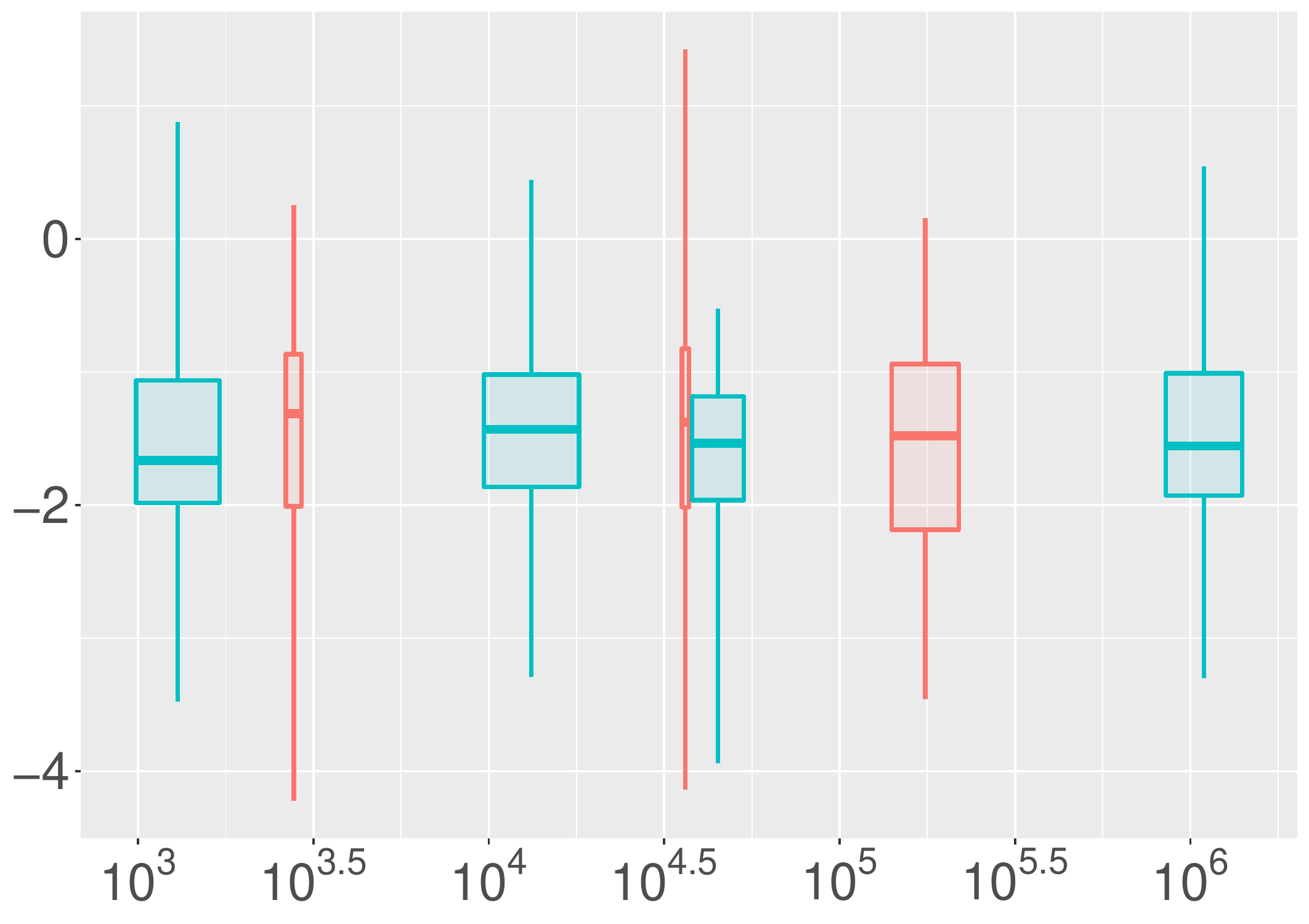}};
\node[above=of img3, yshift = -0.9cm] {$(1,1)$};
\node[above=of img4, yshift = -0.9cm] {$(8,8)$};
\node[left=of img3, node distance = 0, rotate = 90, anchor = center, yshift = -0.5cm] {$d=16\times16$};
\node[below=of img3, yshift = 0.8cm] {Runtime / s};
\node[below=of img4, yshift = 0.8cm] {Runtime / s};
\node[below=of img3, node distance = 0, xshift = 3cm, yshift = 0.5cm]{\includegraphics[width = 0.6\textwidth]{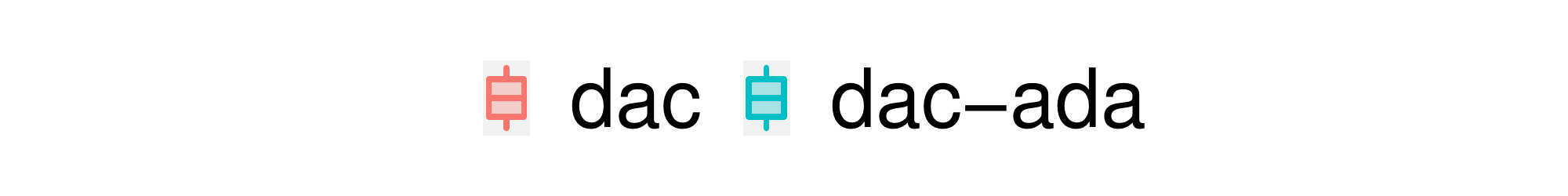}};
\end{tikzpicture}
\caption{Filtering mean estimates for a corner node and a node in the middle of the grid for a $8\times8$ and a $16\times16$ lattice at time $t=10$. The boxplots from left to right report the distributions over 50 repetitions for $N=100, 500, 1000$ and $5000$. The results for the non-adaptive version of DaC are not included for $N=5000$ due to the excessive cost. The reference lines for the $8\times 8$ grid show the average value of the filtering mean estimate and the interquartile range obtained with 50 repetitions of a bootstrap PF with $N=10^5$ particles.
}
\label{fig:spatial}
\end{figure}

To decompose the 2D lattice into a binary tree we use the decomposition described in \citet[Section 5.1]{lindsten2017divide}, which recursively connects the vertices first horizontally and then vertically.
To evaluate the performances of the algorithm we consider the filtering means obtained with 50 repetitions of DaC-SMC on a $8\times8$ and a $16\times16$ grid for $N=100, 500, 1000$ and $5000$. To show how the standard bootstrap particle filter struggles with higher dimensional problems we run a bootstrap PF with $N=10^5$ particles for the $8\times 8$ grid.
Figure~\ref{fig:spatial} reports the filtering means for a corner node and an interior node of the lattice. Both DaC approaches are in agreement, however the adaptive version of DaC seems to provide slightly less variable results. The behaviour for different nodes is similar. 
As observed for the linear Gaussian model, DaC with adaptive lightweight mixture resampling has lower cost than the non-adaptive counterpart and remains feasible for large $N$ (e.g. $N=5000$).

Unsurprisingly, the bootstrap PF struggles to recover the filtering means and provides high variance estimates for node $(1, 1)$ while collapses completely for node $(8, 6)$ failing to recover the filtering mean.

The size of the boxplots in Figure~\ref{fig:spatial} gives an indication of the variance of the estimator provided by DaC, for small $N$ the decay in variance seems to be more pronounced (at least for the adaptive version of DaC) than for large $N$, this is consistent with the decay of the standard deviation which would be expected from the variance expansions in \cite{Kuntz2021} where for small $N$ the higher order contributions to the variance are not yet negligible  (we anticipate that a central limit theorem could be obtained by combining the results of \cite{Kuntz2021} with those for marginal PFs).

\section{Discussion}

We introduced a novel sequential Monte Carlo algorithm, combining ideas from marginal PFs and divide-and-conquer SMC to extend the latter to the filtering context.
This algorithm is based on a novel space decomposition for high dimensional SSM which allows to recursively merge low dimensional marginals of the filtering distribution to obtain the full filtering distribution, taking into account the mismatch between product of marginals and joint distributions using importance sampling. In principle, the DaC-SMC approach is amenable to distributed implementation, although the marginalization technique employed herein would necessitate significant communication from the node which computes the overall filtering distribution at time $t-1$ to all nodes involved in computing at time $t$ and we have not explored that direction here.
 
In contrast with Nested SMC and space-time PFs, the DaC-SMC approach to filtering can be applied when the marginals of the joint density~\eqref{eq:joint} are not available analytically.
The computational cost of this new approach grows polynomially with the number of particles $N$, however, this cost can be reduced 
exploiting GPU routines to reduce the cost of computing the weights as discussed in Section~\ref{sec:cost}.

The experiments in Section~\ref{sec:lgssm} show that DaC-SMC achieves comparable performances of NSMC and STPF with a runtime that remains competitive even for large $d$ (but small $N$), contrary to STPFs. In addition, DaC-SMC can be applied to filtering problems which do not allow for factorization as shown in Section~\ref{sec:ng} and Appendix~\ref{app:expe}.
The variance decay in Figure~\ref{fig:spatial} and Figure~\ref{fig:app_lgssm} in Appendix~\ref{app:expegaussian} suggest that this extended DaC-SMC achieves the same convergences rates as DaC-SMC \citep{Kuntz2021} for sufficiently large $N$ and we anticipate that techniques used to analyze the marginal particle filter could be combined with those in order to provide formal convergence results for the method developed herein.
The adaptive lightweight mixture resampling discussed in Section~\ref{sec:light} is a promising route to further reduce the computational cost of DaC-SMC for filtering, however, as the experiments in Section~\ref{sec:expe} and Appendix~\ref{app:resampling} suggest, selecting the value at which the target $\ess$ should be set to obtain the best trade-off between computational cost and accuracy is likely to be problem dependent and raises interesting theoretical questions that we leave for future work.

For challenging problems it is likely that tempering and MCMC kernels would be required to give good performance. As discussed in \cite{johansen2015blocks} and \cite{Guarniero2017} the smoothing and filtering distributions (i.e. $p(x_{1:t}\vert y_{1:t})$ and $p(x_{t}\vert y_{1:t})$, respectively) have significantly different support in the presence of informative observations, especially in high dimensional settings and so we expect that including the influence of future observations in the proposals and targets in Section~\ref{sec:proposal} (as in lookahead methods, e.g., \citet{Lin2013,Guarniero2017,ruzayqat2021lagged}) would lead to considerable improvements in the accuracy of the estimates and might ultimately be essential in the development of good general purpose filters for high dimensional problems.

This work focuses on obtaining approximations of the filtering distribution for high dimensional SSM. In recent years there has been a lot of interest in obtaining approximations of the smoothing distribution which is a necessary component of parameter estimation algorithms (e.g., \cite{finke2017approximate, Guarniero2017}); we anticipate that the DaC-SMC approach to filtering could be extended to tackle smoothing and parameter estimation dealing with the marginalization in~\eqref{eq:gamma}. In principle, algorithms which directly approximate only marginals of smoothing distributions can be adapted to these settings (see, e.g. \cite{gerber2017convergence}). 
We leave this for future work.
\section*{Supplementary Materials}

Supplementary materials contains details of the tempering approach, additional details and results on the experiments. An \texttt{R} package reproducing the experiments is available at \url{https://github.com/FrancescaCrucinio/Dac4filtering}.
\par
\section*{Acknowledgements}	
FRC and AMJ acknowledge support from the EPSRC (grant \#  EP/R034710/1). AMJ acknowledges further support from the  EPSRC (grant \# EP/T004134/1) and the Lloyd's Register Foundation Programme on Data-Centric Engineering at the Alan Turing Institute. 
For the purpose of open access, the author has applied a Creative Commons Attribution (CC BY) licence to any Author Accepted Manuscript version arising from this submission.
\par
	
\if\version1 
\bibhang=1.7pc
\bibsep=2pt
\fontsize{9}{14pt plus.8pt minus .6pt}\selectfont
\renewcommand\bibname{\large \bf References}
\expandafter\ifx\csname
natexlab\endcsname\relax\def\natexlab#1{#1}\fi
\expandafter\ifx\csname url\endcsname\relax
  \def\url#1{\texttt{#1}}\fi
\expandafter\ifx\csname urlprefix\endcsname\relax\def\urlprefix{URL}\fi
\fi
\bibliographystyle{chicago}      
\bibliography{dac4filtering_biblio}   

\if\version1 
\vskip .65cm
\noindent
Department of Statistics, University of Warwick
\vskip 2pt
\noindent
E-mail: francesca.crucinio@warwick.ac.uk
\vskip 2pt

\noindent
Department of Statistics, University of Warwick
\vskip 2pt
\noindent
E-mail: a.m.johansen@warwick.ac.uk
\fi

\if\version0

\newpage
\appendix

\section{Comparison of Resampling Strategies}
\label{app:resampling}

The divide and conquer SMC algorithm described in Section~\ref{sec:dac} merges the particle populations of each node's children using mixture resampling.
We compare the performances of lightweight mixture resampling with its adaptive version described in Section~\ref{sec:light}, those of full-cost mixture resampling, in which all possible $N^2$ permutations of the particles are created, and those of a linear cost version of DaC-SMC in which no mixing weights are used \citep{lindsten2017divide}.

To compare the four resampling schemes we consider the mean squared error ($\mse$) for each component $i$ using the example of Section~\ref{sec:lgssm},
\begin{align}
\label{eq:mse}
\mse(x_{t}(i)):=\mathbb{E}\left[(\bar{x}_{t}(i)-\mu_{t, i})^2\right],
\end{align}
where $\bar{x}_{t}(i)$ denotes the estimate of the mean of component $i$ at time $t$ and $\mu_{t, i}$ denotes the true mean of $x_{t}(i)\vert y_{1:t}$ obtained from the Kalman filter. To approximate~\eqref{eq:mse} we consider an empirical average over 50 repetitions.

Figure~\ref{fig:lc_vs_mix} shows the average (over component) $\mse$ of the estimates obtained for $d=128$ and $t=10$ time steps as a function of runtime.
The linear cost version of DaC-SMC has the smallest runtime, however, the results in terms of $\mse$ are the worst among the four algorithms. In fact, this linear cost version does not use mixture weights at the point of selection, and therefore does not take into account the mismatch between $\gamma_{t, \mathcal{C}_u} $ and $ \gamma_{t, u}$ when resampling (this is corrected for with a subsequent importance reweighting). The full mixture resampling has a higher cost, and becomes unmanageable for large $N$ (it is, in fact, not included for $N= 1000,5000$); the non-adaptive version of lightweight mixture resampling also becomes too expensive for large $N$ although the increase in cost is less steep than that of the full mixture resampling (the non-adaptive lightweight mixture resampling has manageable cost for $N=1000$ but it is not included for $N=5000$).
The $\mse$ achieved by the non-adaptive lightweight mixture resampling is equivalent to that of the full mixture resampling, at a considerably lower cost. For small $N$, the adaptive lightweight mixture resampling also gives comparable results, but, as $N$ increases, the $\mse$ stops improving and eventually settles around $0.02$. 
This is likely due to the fact that, as it is the case of Figure~\ref{fig:m_adaptive}, sometimes the target $\ess$ is not reached at level 1 and more permutation would be needed to obtain a good importance sampling proposal.
One could then consider to add tempering steps to obtain better samples or, alternatively, to increase the target $\ess$. By how much the target $\ess$ should be increased is a challenging question
which have not investigated this further in this work.

\begin{figure}
\centering
\begin{tikzpicture}[every node/.append style={font=\normalsize}]
\node (img1) {\includegraphics[width = 0.8\textwidth]{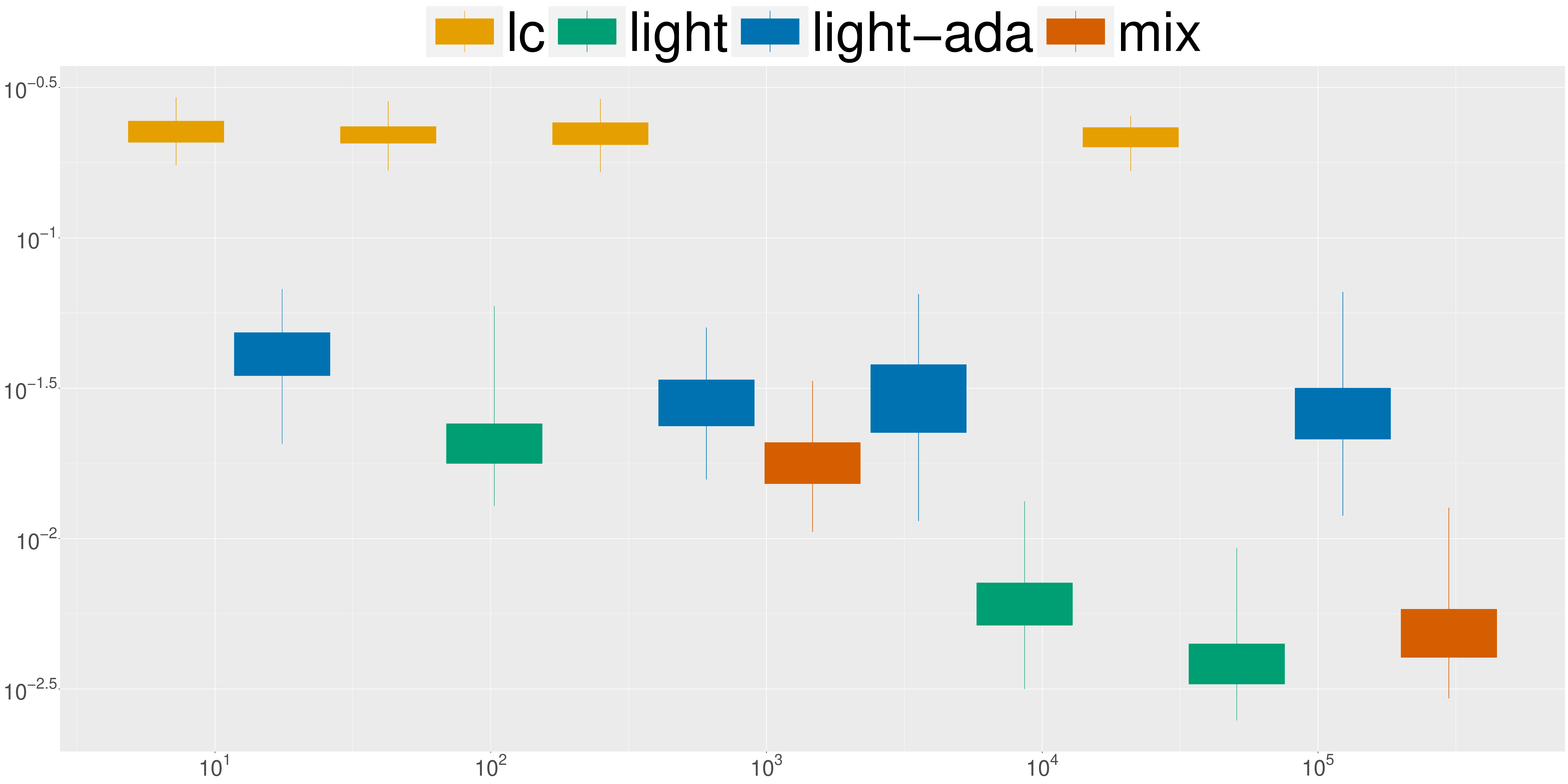}};
\node[below=of img1, node distance = 0, yshift = 1cm] {Runtime / s};
\end{tikzpicture}
\caption{Average (over dimensions) $\mse$ at $t= 10$ over 50 runs for $d=128$ as function of runtime for the linear cost version of DaC in \cite{lindsten2017divide} and Algorithm~\ref{alg:dac_filter} with lightweight mixture resampling, adaptive lightweight mixture resampling and full mixture resampling. The boxes, from left to right, correspond to increasing number of particles, $N=100, 500, 1000, 5000$. Due to the excessive cost, the results for mixture resampling are not reported for $N=1000, 5000$ and those for lightweight mixture resampling with no adaptation are not reported for $N= 5000$.}
\label{fig:lc_vs_mix}
\end{figure}

\section{Tempering Strategy}
\label{app:tempering}
If the product of the marginals over the two child nodes merged via mixture resampling provides a poor approximation for $\gamma_{t,u}$ (which in some circumstances could be detected, for example, by Algorithm~\ref{alg:adaptive_light} failing achieve the target $\ess$), one could expect to alleviate this mismatch via a tempering strategy. Here, we describe an adaptive tempering strategy of the sort described in \citet[Section 4.2]{lindsten2017divide} adapted to our context.

We define the following sequence of targets $\hat{\pi}_{\alpha_{u,j}}=\hat{\gamma}_{\alpha_{u,j}}/\hat{\gamma}_{\alpha_{u,j}}(1)$ where
\begin{align*}
\hat{\gamma}_{\alpha_{u,j}}(z_{t,u, j}) = \left[\gamma_{t, \mathcal{C}_u}(z_{t,u, j})\right]^{1-\alpha_{u,j}} \gamma_{t,u}(z_{t,u, j})^{\alpha_{u,j}}
\end{align*}
for $z_{t,u, j}\in\real^{\vert \mathcal{V}_u\vert}$ and $\alpha_{u,j}\in[0, 1]$, which interpolates from the proposal $\gamma_{t, \mathcal{C}_u}$ to the target at node $u$, $\gamma_{t,u}$.
In a standard approach to tempering we would start from $\alpha_{u, j}=0$ and select the next value adaptively (see e.g. \citet{wang2020annealed, zhou2016toward}). However, Algorithm~\ref{alg:adaptive_light} corresponds to a first tempering step which moves the proposal $\gamma_{t, \mathcal{C}_u}$ closer to the target $\gamma_{t,u}$; therefore, 
instead of starting the tempering schedule at $\alpha_{u, j}=0$, we identify the value of $\alpha^\star$ which corresponds to the intermediate tempered target obtained after mixture resampling and carry on with tempering from there until $\alpha_{u, j}=1$.
To identify the value of $\alpha$ we consider the $\ess$
\begin{align*}
\ess(\alpha)&:= \left(\sum_{n=1}^{N} (\tilde{w}_{t,u}^{n})^{\alpha}\right)^2/\sum_{n=1}^{N} (\tilde{w}_{t,u}^{n})^{2\alpha},
\end{align*}
and solve $\ess(\alpha) = \ess^\star$, where $\ess^\star$ is the target $\ess$ in Section~\ref{sec:light}.
The equally weighted particles after mixture resampling, $\{z_{t,\mathcal{C}_u}, \omega_{u,j}=1\}_{n=1}^N$, approximate $\hat{\pi}_{\alpha^\star}$.
The following values of the tempering sequence are chosen adaptively as described in, e.g., \citet[Section 3.3.2 and Algorithm 4]{zhou2016toward} and \citet[Algorithm 4]{wang2020annealed}, using the conditional $\ess$
\begin{align}
\label{eq:ess_tempering}
\cess_j(\alpha)& := N\left(\sum_{n=1}^{N} \bar{\omega}_{u, j}^{n}(\tilde{w}_{t,u}^{n})^{\alpha - \alpha_{u, j-1}}\right)^2/\sum_{n=1}^{N} \bar{\omega}_{u, j}^{n}(\tilde{w}_{t,u}^{n})^{2(\alpha-\alpha_{u,j-1})},
\end{align}
where $\bar{\omega}_{u, j}^{n}\propto\omega_{u, j}^{n}$.
Given a decay threshold $\beta>0$, we find the next value $\alpha_{u,j}$ solving $\cess_j(\alpha)=\beta$, then we update the weights, perform a resampling step if necessary, and rejuvenate the particles using a $\hat{\pi}_{\alpha_{u,j}}$-invariant kernel $K'_{\alpha_{u, j}}$.
This process is repeated until $\alpha_{u,j}=1$.
The resulting tempering strategy is summarized in  Algorithm~\ref{alg:tempering}.

\begin{algorithm}[th]
\begin{algorithmic}[1]
\STATE{\textit{Initialize:} set $\alpha_{u,1}$ to be the solution of $\ess(\alpha) = \ess^{\star}$ and set $\omega_{u, 1}^{n} = 1$ for $n\leq N$.}
\STATE{\textit{Set:} $j=1$.}
\WHILE{$\alpha_{u, j}<1$}
\STATE{\textit{Adapt:} set $j=j+1$ and find $\alpha_{u, j}$ solving $\cess_j(\alpha) = \beta$ for $\alpha\in(\alpha_{u,j-1}, 1)$ using bisection.}
\STATE{\textit{Reweight:} compute the weights $\omega_{u, j}^{n} =\omega_{u, j-1}^{n, N} (\tilde{w}_{t, u}^{n})^{\alpha_{u, j}-\alpha_{u, j-1}} $ for $n\leq N$ and compute the $\ess$.}
\IF{$\ess<N/2$}
\STATE{\textit{Resample:} draw new $ z_{t,u, j-1}^{n,N}$ independently with weights $\omega_{u, j}^{n}$ and update $\omega_{u, j}^{n}=1$ for $n\leq N$.}
\ENDIF
\STATE{\textit{Propose:} draw $z_{t,u, j}^{n}\sim K'_{\alpha_{u, j}}(\cdot, z_{t,u, j-1}^{n})$ from a $\hat{\pi}_{\alpha_{u, j}}$ invariant kernel for $n\leq N$.}
\ENDWHILE
\end{algorithmic}
\caption{Tempering.}\label{alg:tempering}
\end{algorithm}

\section{Further Details on the Experiments}
\label{app:details}
We collect here further details on the models considered in Section~\ref{sec:expe}.

\subsection{Simple Linear Gaussian Model}\label{app:detailsgaussian}
We consider the simple linear Gaussian model used in the experiments in \citet[Section 6.1]{naesseth2015nested}: 
\begin{align*}
X_1&\sim \N_d (0, \Id_d)\\
X_t &= 0.5AX_{t-1} + U_t, \qquad U_t\sim \N_d(0, \Sigma)\\
Y_t &= X_t + V_t,\qquad V_t \sim \N_d (0, \sigma_y^2 \Id_d),
\end{align*}
with $A =  A_1A_2^{-1}$ where
\begin{align*}
&A_1=\begin{pmatrix}
\tau+ \lambda & 0 & 0 &\dots &\dots & 0 & 0\\
0 & \tau & 0 & 0 & \dots & 0 & 0\\
0 & \ddots & \ddots & \ddots & \ddots & 0 & 0\\
\vdots & \ddots & \ddots & \ddots & \ddots & \vdots & \vdots\\
\vdots & \ddots & \ddots & \ddots & \ddots & \vdots & \vdots\\
0 & 0 & 0 & 0 & 0 & \tau  & 0\\
0 & 0 & 0 & 0 & 0 &0 & \tau  
  \end{pmatrix},
\end{align*}
\begin{align*}
&A_2=\begin{pmatrix}
\tau+ \lambda & 0 & 0 &\dots &\dots & 0 & 0\\
- \lambda & \tau + \lambda & 0 & 0 & \dots & 0 & 0\\
0 & \ddots & \ddots & \ddots & \ddots & 0 & 0\\
\vdots & \ddots & \ddots & \ddots & \ddots & \vdots & \vdots\\
\vdots & \ddots & \ddots & \ddots & \ddots & \vdots & \vdots\\
0 & 0 & 0 & 0 & - \lambda & \tau  + \lambda & 0\\
0 & 0 & 0 & 0 & 0 &- \lambda & \tau  + \lambda
\end{pmatrix}^{-1},
\end{align*}
and
\begin{align*}
\Sigma^{-1} = A_2^T\begin{pmatrix}
\tau & 0 & 0 &\dots &\dots & 0 & 0\\
0 & \tau + \lambda & 0 & 0 & \dots & 0 & 0\\
0 & \ddots & \ddots & \ddots & \ddots & 0 & 0\\
\vdots & \ddots & \ddots & \ddots & \ddots & \vdots & \vdots\\
\vdots & \ddots & \ddots & \ddots & \ddots & \vdots & \vdots\\
0 & 0 & 0 & 0 & 0 & \tau  + \lambda & 0\\
0 & 0 & 0 & 0 & 0 &0 & \tau  + \lambda
\end{pmatrix}A_2,
\end{align*}
with $\tau=\lambda = 1$ and $\sigma^2_y = 0.5^2$.

To derive the mixture weights for our divide-and-conquer approach we write the joint density of $( X_{1:t}, Y_{1:t})$:
\begin{align*}
p(x_ {1:t}, y_{1:t}) &\propto \prod_{k=1}^t\left[ \exp\left(-\frac{1}{2}(x_k - 0.5Ax_{k-1})^T\Sigma^{-1}(x_k - 0.5Ax_{k-1})\right)\right.\\
&\qquad\times \left.\prod_{i=1}^d \N(y_{k}(i); x_{k}(i), \sigma_y^2)\right].
\end{align*}
The functions $f_{t,u}, g_{t, u}$ in the definition of $\gamma_{t,u}$ in~\eqref{eq:gamma} are
\begin{align*}
&f_{t,u}(x_{t-1}, x_t(\mathcal{V}_{u})) \propto \\
&\qquad\exp\left(-\frac{1}{2}(x_t(\mathcal{V}_{u}) - 0.5A_{\mid \mathcal{V}_u}x_{t-1}(\mathcal{V}_{u}))^T\Sigma_{\mid \mathcal{V}_u}^{-1}(x_t(\mathcal{V}_{u}) - 0.5A_{\mid \mathcal{V}_u}x_{t-1}(\mathcal{V}_{u}))\right),\\
&g_{t, u}(x_t(\mathcal{V}_{u}), (y_t(i))_{i\in\mathcal{V}_{u}})\propto\prod_{i \in\mathcal{V}_u} \N(y_{t}(i); x_{t}(i), \sigma_y^2),
\end{align*}
where we denote $x_t(\mathcal{V}_{u})=(x_t(i))_{i\in\mathcal{V}_{u}}$ and $A_{\mid \mathcal{V}_u}, \Sigma_{\mid \mathcal{V}_u}^{-1}$ are the restrictions of $A, \Sigma^{-1}$, respectively, to $\mathcal{V}_u$, i.e. $A_{\mid \mathcal{V}_u}$ is the matrix obtained discarding the elements of $A$ corresponding to components which are not in $ \mathcal{V}_u$, and similarly for $\Sigma_{\mid \mathcal{V}_u}^{-1}$.

Expanding the quadratic form in the display above, we find that we can decouple the dependence on the current time step from that on the past and decompose
\begin{align*}
f_{t,u}(x_{t-1}, (x_t(i))_{i\in\mathcal{V}_{u}}) & = f_{t,u}^{(1)}(x_{t-1}, (x_t(i))_{i\in\mathcal{V}_{u}})\prod_{i\in\mathcal{V}_u, i\neq j_u^{1}}\tilde{f}(x_t(i-1), x_t(i)),
\end{align*}
with
\begin{align*}
f_{t,u}^{(1)}(x_{t-1}, (x_t(i))_{i\in\mathcal{V}_{u}}) &\propto\prod_{i\in\mathcal{V}_u, i\neq j_u^{1}} \left[\exp\left(-\frac{\tau+\lambda}{2}\left(x_{t}(i)- 0.5\frac{\tau}{\tau+\lambda}x_{t-1}(i)\right)^2\right)\right.\\
&\qquad\left.\times \exp\left(-\frac{1}{2} \frac{\lambda\tau}{\tau+\lambda} x_t(i-1)x_{t-1}(i)\right)\right]\\
&\times \begin{cases}
\exp\left(-\frac{\tau+\lambda}{2}\left(x_{t}(j_u^1)- 0.5\frac{\tau}{\tau+\lambda}x_{t-1}(j_u^1)\right)^2\right)\qquad\textrm{if } j_u^1\neq 1\\
\exp\left(-\frac{\tau}{2}(x_{t}(1)- 0.5x_{t-1}(1))^2\right)\qquad\textrm{if } j_u^1= 1
\end{cases},
\end{align*}
where $j_u^{1}$ denotes the first index associated with node $u$, and
\begin{align*}
\tilde{f}(s_1, s_2) := \exp\left(-\frac{1}{2}\left[\frac{\lambda^2}{\tau+\lambda}s_1^2-2\lambda s_1s_2\right] \right).
\end{align*}
The mixture weights are then given by 
\begin{align*}
    m_{t, u}(z_{t,\mathcal{C}_u}) &\propto \tilde{f}\left(z_{t, \ell(u)}(i_{\ell(u)}^{n_{\ell(u)}}), z_{t,r(u)}(i_{r(u)}^{1})\right)\\
    &\times \frac{\sum_{n=1}^N f_{t,u}^{(1)}(z_{t-1, \mathfrak{R}}^n, z_{t, \mathcal{C}_u})}{\sum_{n=1}^N f_{t,\ell(u)}^{(1)}(z_{t-1, \mathfrak{R}}^n, z_{t, \ell(u)})\sum_{n=1}^N f_{t,r(u)}^{(1)}(z_{t-1, \mathfrak{R}}^n, z_{t, r(u)})},
\end{align*}
where we recall that $z_{t,u}$ in~\eqref{eq:gamma} corresponds to $z_{t, u}=(x_t(i))_{i\in\mathcal{V}_{u}}$, $i_{\ell(u)}^{n_{\ell(u)}}$ is the last index associated with node $\ell(u)$ and $i_{r(u)}^{1}$ is the first index associated with $r(u)$. 

\subsection{Spatial Model}

The joint density of $( X_{1:t}, Y_{1:t})$ is
\begin{align*}
p(x_ {1:t}, y_{1:t}) &\propto \prod_{k=1}^t\left(\prod_{v\in V} \N\left(x_k(v); x_{k-1}(v), \sigma_x^2\right)\right.\\
&\left.\times\left[1+\nu^{-1}\sum_{v\in V}\left((y_k(v)-x_k(v))\sum_{j: D(v, j) \leq r_y}\tau^{D(v, j)}(y_k(j)-x_k(j))\right)\right]^{-(\nu+\vert V\vert)/2}\right)
\end{align*} 
with initial distribution $X_1\sim \prod_{v\in V} \N\left(x_k(v); 0, \sigma_x^2\right)$.
The functions $f_{t,u}, g_{t, u}$ in the definition of $\gamma_{t,u}$ in~\eqref{eq:gamma} are
\begin{align*}
&f_{t,u}(x_{t-1}, (x_t(i))_{i\in\mathcal{V}_{u}})\propto\prod_{v\in \mathcal{V}_u} \N\left(x_k(v); x_{k-1}(v), \sigma_x^2\right)\\
&g_{t,u}((x_t(i))_{i\in\mathcal{V}_{u}}, (y_t(i))_{i\in\mathcal{V}_{u}}) \propto\\
&\qquad \left[1+\nu^{-1}\sum_{v\in \mathcal{V}_u}\left((y_k(v)-x_k(v))\sum_{j: D(v, j) \leq r_y, j\in \mathcal{V}_u}\tau^{D(v, j)}(y_k(j)-x_k(j))\right)\right]^{-(\nu+\vert \mathcal{V}_u\vert)/2},
\end{align*}
where we recall that $z_{t,u}$ in~\eqref{eq:gamma} corresponds to $z_{t, u}=(x_t(i))_{i\in\mathcal{V}_{u}}$.
From the above, we obtain that the mixture weights~\eqref{eq:w_mix} are given by $m_{t, u}(z_{t,\mathcal{C}_u}) = R^f_{t,u}(z_{t,\mathcal{C}_u})R^g_{t,u}(z_{t,\mathcal{C}_u})$, where
\begin{align*}
R^f_{t,u}(z_{t,\mathcal{C}_u}) &=\frac{N^{-1}\sum_{n=1}^N f_{t,u}(z_{t-1, \mathfrak{R}}^n, z_{t, \mathcal{C}_u})}{ N^{-1}\sum_{n=1}^Nf_{t,\ell(u)}(z_{t-1, \mathfrak{R}}^n, z_{t,\ell(u)}) N^{-1}\sum_{n=1}^N f_{t,r(u)}(z_{t-1, \mathfrak{R}}^n, z_{t,r(u)})}\\
&\propto\frac{\sum_{n=1}^N\prod_{v\in \mathcal{V}_u} \N\left(z_{t,\mathcal{C}_u}(v); z_{t-1, \mathfrak{R}}^n(v), \sigma_x^2\right)}{\sum_{n_1=1}^N\prod_{v\in \mathcal{V}_{\ell(u)}} \N\left(z_{t,\ell(u)}(v); z_{t-1, \mathfrak{R}}^{n_1}(v), \sigma_x^2\right)\sum_{n_2=1}^N\prod_{v\in r(u)} \N\left(z_{t,r(u)}(v); z_{t-1, \mathfrak{R}}^{n_2}(v), \sigma_x^2\right)}
\end{align*}
and
\begin{align*}
R^g_{t, u}(z_{t,\mathcal{C}_u}) &=\frac{g_{t,u}(z_{t, \mathcal{C}_u}, (y_t(i))_{i\in\mathcal{V}_u})}{g_{t,\ell(u)}(z_{t,\ell(u)}, (y_t(i))_{i\in\mathcal{V}_{\ell(u)}})g_{t,r(u)}(z_{t,r(u)}, (y_t(i))_{i\in\mathcal{V}_{r(u)}})}\\
&   \propto \left[1+\nu^{-1}\sum_{v\in \mathcal{V}_u}\left((y_t(v)-z_{t,\mathcal{C}_u}(v))\sum_{\substack{j: D(v, j) \leq r_y\\ j\in\mathcal{V}_u}}\tau^{D(v, j)}(y_t(j)-z_{t,\mathcal{C}_u}(j))\right)\right]^{-\frac{\nu+\vert \mathcal{V}_{u}\vert}{2}}\\
    &\times \left[1+\nu^{-1}\sum_{v\in \mathcal{V}_{\ell(u)}}\left((y_t(v)-z_{t,\ell(u)}(v))\sum_{\substack{j: D(v, j) \leq r_y\\ j\in\mathcal{V}_{\ell(u)}}}\tau^{D(v, j)}(y_t(j)-z_{t,\ell(u)}(j))\right)\right]^{\frac{\nu+\vert \mathcal{V}_{\ell(u)}\vert}{2}}\\
    &\times\left[1+\nu^{-1}\sum_{v\in \mathcal{V}_{r(u)}}\left((y_t(v)-z_{t,r(u)}(v))\sum_{\substack{j: D(v, j) \leq r_y\\ j\in\mathcal{V}_{r(u)}}}\tau^{D(v, j)}(y_t(j)-z_{t,r(u)}(j))\right)\right]^{\frac{\nu+\vert \mathcal{V}_{r(u)}\vert}{2}}.
\end{align*}

\section{Additional Results for the Experiments}
\label{app:expe}
\subsection{Simple Linear Gaussian Model}\label{app:expegaussian}

To further characterize the quality of the estimates obtained with Algorithm~\ref{alg:dac_filter}, we consider the relative mean squared error ($\rmse$) for component $i$ at time $t$,
\begin{align*}
\rmse(x_{t}(i)):=\frac{\mathbb{E}\left[(\bar{x}_{t}(i)-\mu_{t, i})^2\right]}{\sigma^2_{t, i}},
\end{align*}
where where $\bar{x}_{t}(i)$ denotes the estimate of the mean of component $i$ at time $t$ and $\mu_{t, i}, \sigma^2_{t,i}$ denote the true mean and true variance of $x_{t}(i)\vert y_{1:t}$ obtained from the Kalman filter. 
As for~\eqref{eq:mse}, we approximate the $\rmse$ using an empirical average over 50 repetitions.
\begin{figure}
\centering
\begin{tikzpicture}[every node/.append style={font=\normalsize}]
\node (img1) {\includegraphics[width = 0.9\textwidth]{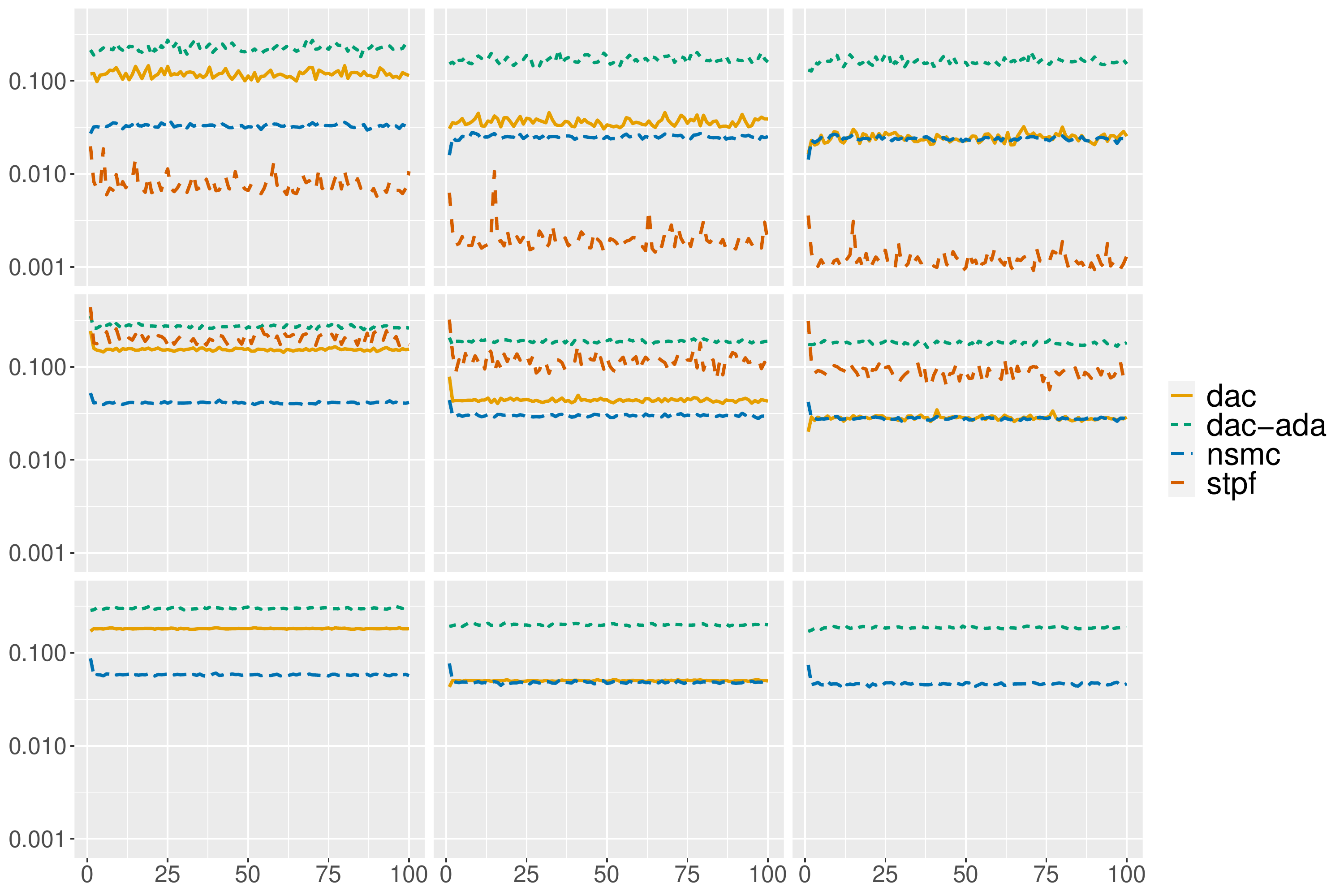}};
\node[above=of img1, node distance = 0, yshift = 1cm, yshift = -2cm, xshift = -4.5cm] {$N=100$};
\node[above=of img1, node distance = 0, yshift = 1cm, yshift = -2cm, xshift = -0.5cm] {$N=500$};
\node[above=of img1, node distance = 0, yshift = 1cm, yshift = -2cm, xshift = 3.5cm] {$N=1000$};
\node[left=of img1, node distance = 0, rotate=90, anchor = center, yshift = -0.8cm, xshift = 3cm] {$d=32$};
\node[left=of img1, node distance = 0, rotate=90, anchor = center, yshift = -0.8cm] {$d=256$};
\node[left=of img1, node distance = 0, rotate=90, anchor = center, yshift = -0.8cm, xshift = -3cm] {$d=2048$};
\node[below=of img1, node distance = 0, yshift = 1cm] {Time step};
\end{tikzpicture}
\caption{Average (over dimension) $\rmse$ for 50 runs of DaC, NSMC and STPF. 
Due to their excessive cost, we do not include the results for STPF with $d=2048$ and those of the non-adaptive version of DaC for $d=2048, N=1000$.}
\label{fig:app_lgssm}
\end{figure}
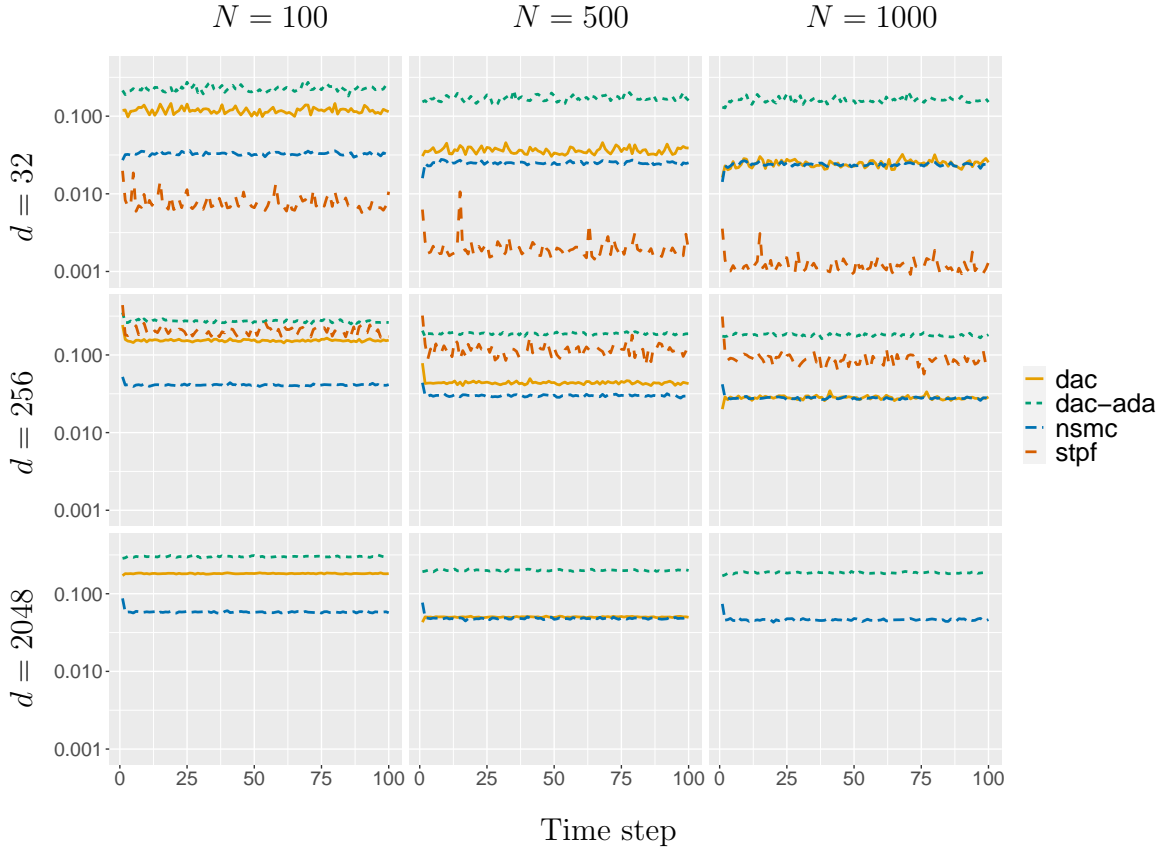

As observed in Section~\ref{sec:lgssm}, for small $d$, STPF performs better than DaC and NSMC (Figure~\ref{fig:app_lgssm} top row), however, as dimension increases, the estimates provided by STPF become poor and their computational cost is prohibitive for large $d$ (Figure~\ref{fig:app_lgssm} bottom row). 
Contrary to the other approaches, NSMC does not seem to significantly improve as $N$ increases.

\subsection{Spatial Model}
To validate the results for the spatial model in Section~\ref{sec:ng} we compare the approximations obtained by Algorithm~\ref{alg:dac_filter} with both adaptive and non-adaptive lightweight mixture resampling with those of the standard bootstrap particle filter (see, e.g., \citet[Chapter 10]{chopin2020}) on a $2\times 2$ grid (Figure~\ref{fig:spatial_app}). Given the low dimensionality of the state space, we expect the bootstrap PF with a large number of particles to provide a good proxy for the filtering distribution $p(x_t | y_{1:t})$. The results for the three algorithms seem to be in good agreement, and the variance of both DaC algorithms with large $N=5\cdot 10^3, 10^4$ is comparable with that of the particle filter with $N=10^5$.
\begin{figure}
\centering
\begin{tikzpicture}[every node/.append style={font=\normalsize}]
\node (img1) {\includegraphics[width = 0.4\textwidth]{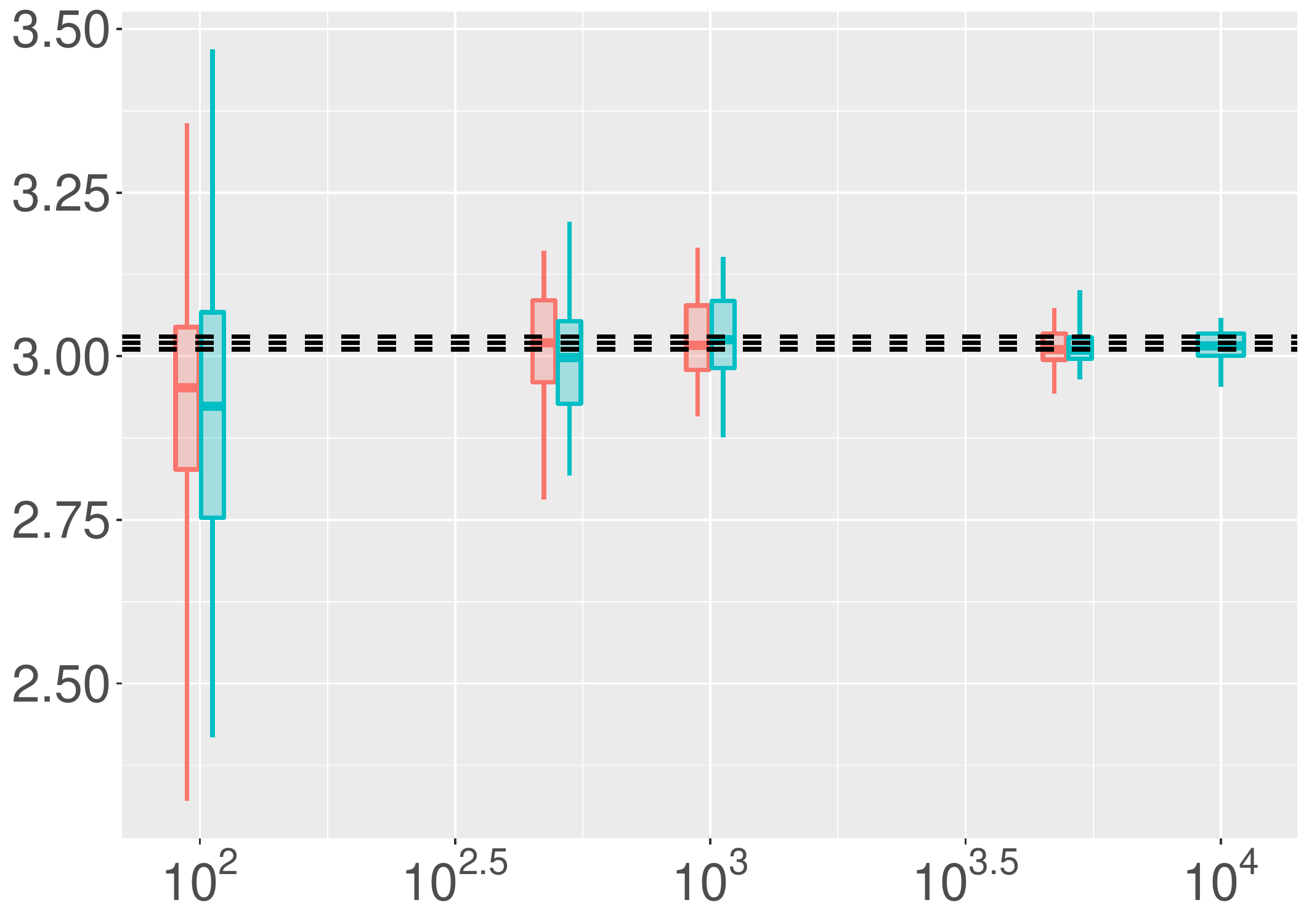}};
\node[right=of img1, xshift = -0.6cm] (img2) {\includegraphics[width = 0.4\textwidth]{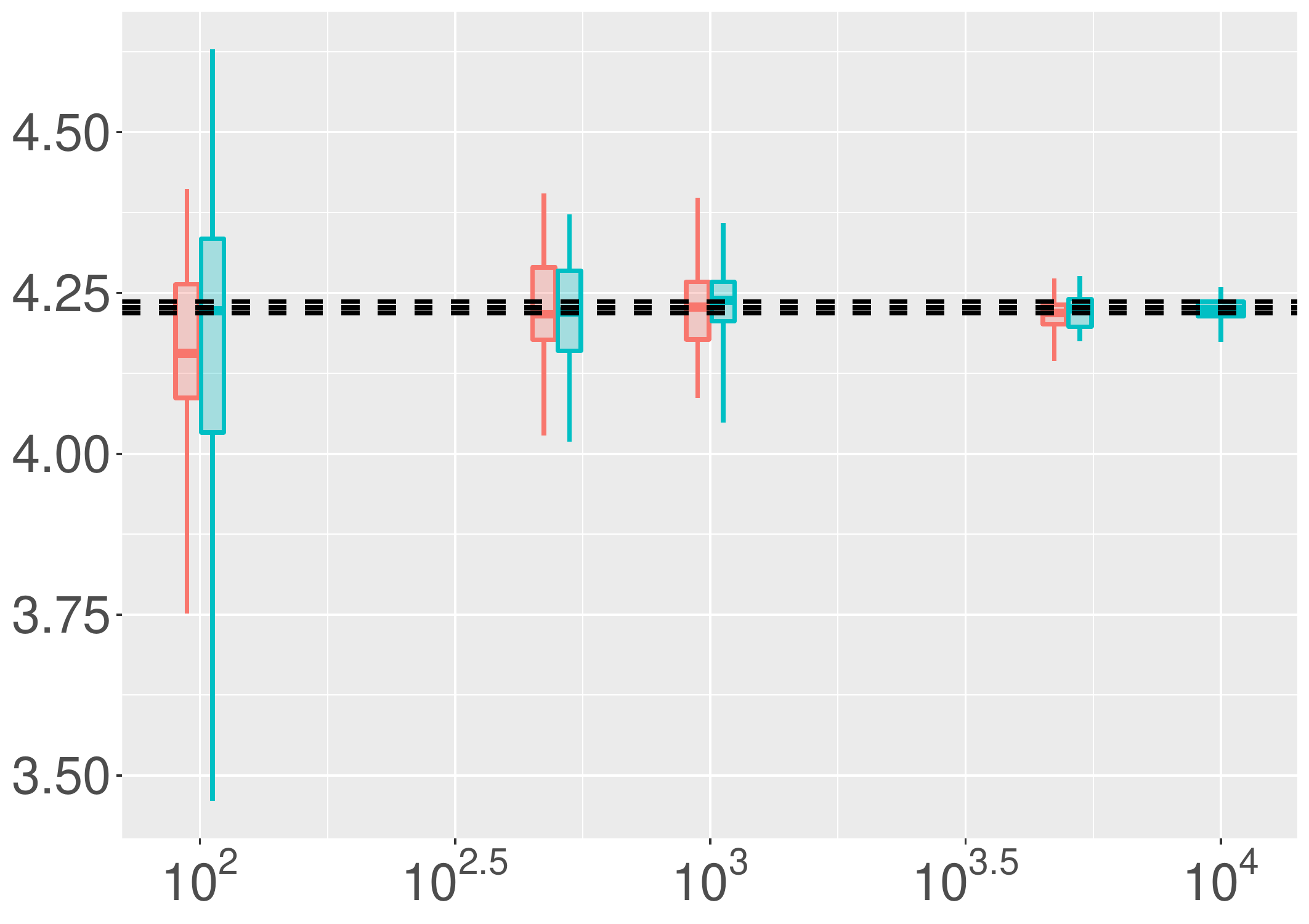}};
\node[below=of img1, yshift = 1cm] (img3) {\includegraphics[width = 0.4\textwidth]{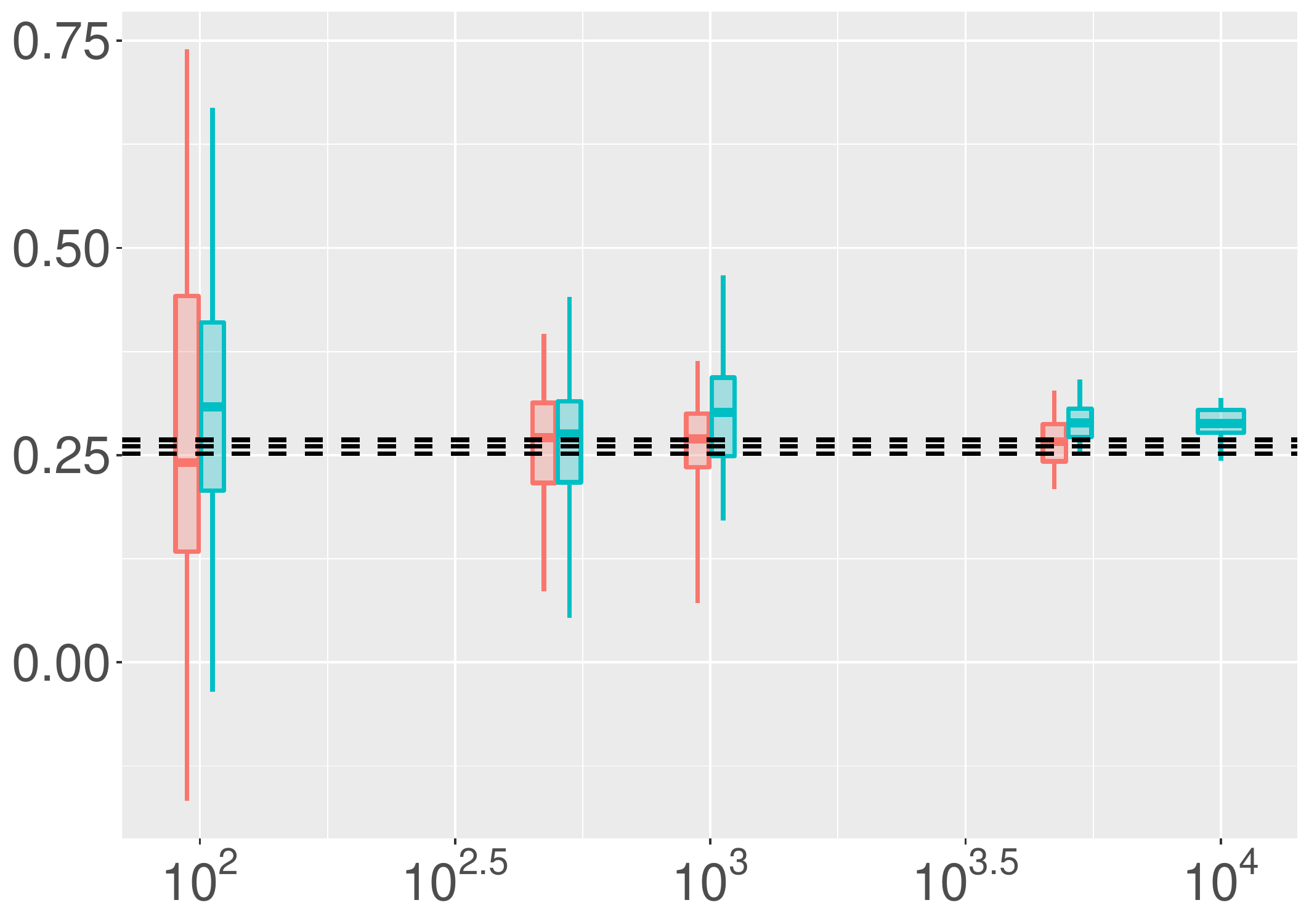}};
\node[right=of img3, xshift = -0.6cm] (img4) {\includegraphics[width = 0.4\textwidth]{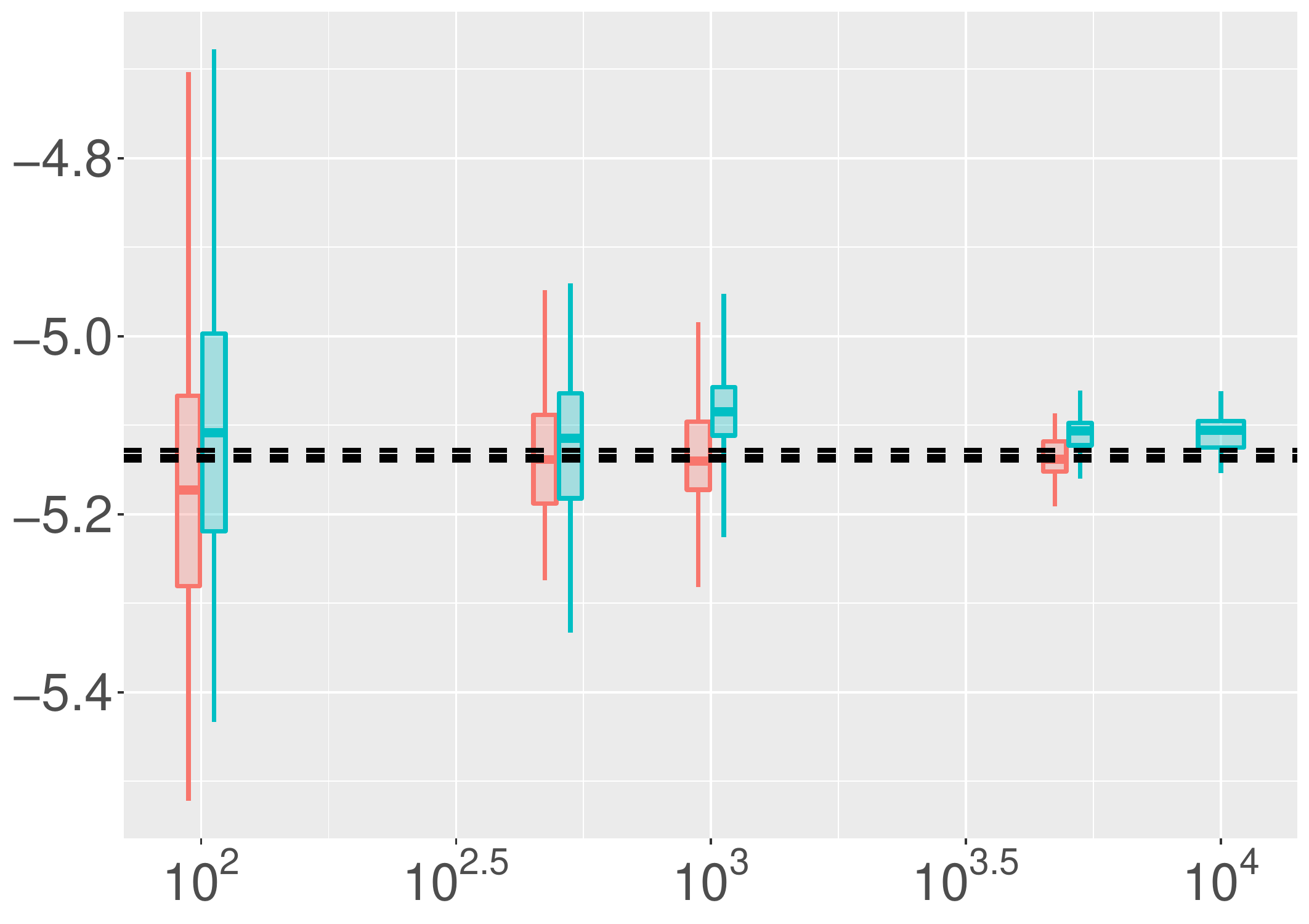}};
\node[below=of img4, xshift = -4cm, yshift = 1cm] (img5) {\includegraphics[width = 0.6\textwidth]{spatial_legend_bpf.pdf}};
\node[above=of img1, yshift = -1cm] {column $1$};
\node[above=of img2, yshift = -1cm] {column $2$};
\node[left=of img1,  rotate = 90, yshift = -0.7cm, xshift = 0.7cm] {row $1$};
\node[left=of img3, rotate = 90, yshift = -0.7cm, xshift = 0.9cm] {row $2$};
\node[below=of img3, yshift = 0.8cm] {$N$};
\node[below=of img4, yshift = 0.8cm] {$N$};
\end{tikzpicture}
\caption{Filtering mean estimates obtained with DaC on a $2\times2$ grid. The reference lines show the average value of the filtering mean estimate and the interquartile range obtained with 50 repetitions of a bootstrap PF with $N=10^5$ particles.
The boxplots from left to right report the distributions over 50 repetitions for $N=100, 500, 1000, 5000$ and $10000$. The results for the non-adaptive version of DaC and $N=10000$ are not included due to the excessive cost.}
\label{fig:spatial_app}
\end{figure}

To show how the increasing dimension causes instability of the simple bootstrap PF we report in Figure~\ref{fig:spatial_app_cod} the filtering mean estimates for two nodes of a $4\times 4$ grid.
Contrary to Figure~\ref{fig:spatial_app}, in which the bootstrap PF with $N=10^5$ particles provides accurate estimates (first quartile, mean and third quartile coincide), when the dimension slightly increases the variance of the particle filter with $N=10^5$ blows up and the recovered estimates are poor. 
On the other hand, both DaC approaches are stable and provide better and better estimates for values of $N$ which are at least 10 times smaller than $N=10^5$.
\begin{figure}
\centering
\begin{tikzpicture}[every node/.append style={font=\normalsize}]
\node (img1) {\includegraphics[width = 0.4\textwidth]{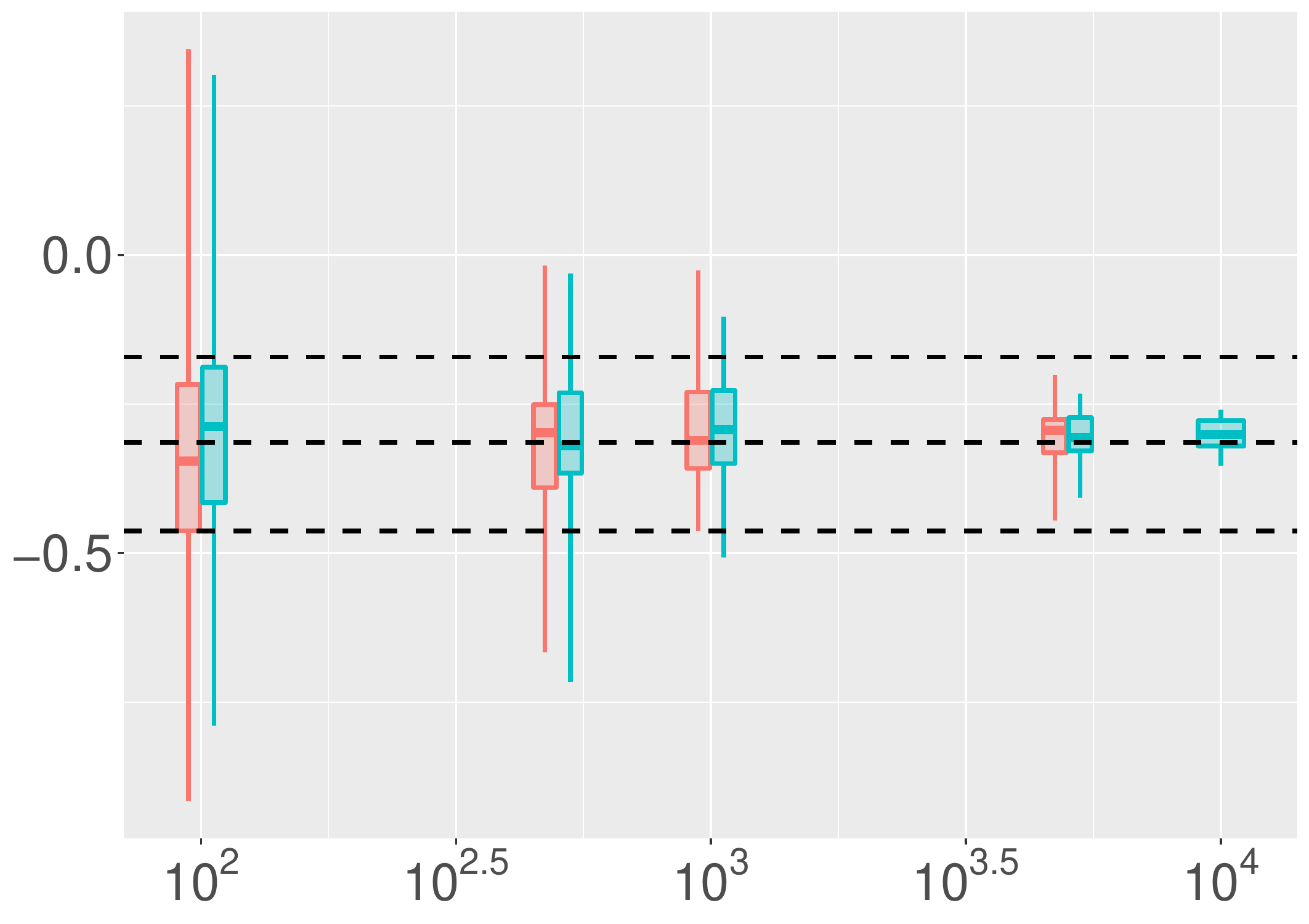}};
\node[right=of img1, xshift = -0.6cm] (img2) {\includegraphics[width = 0.4\textwidth]{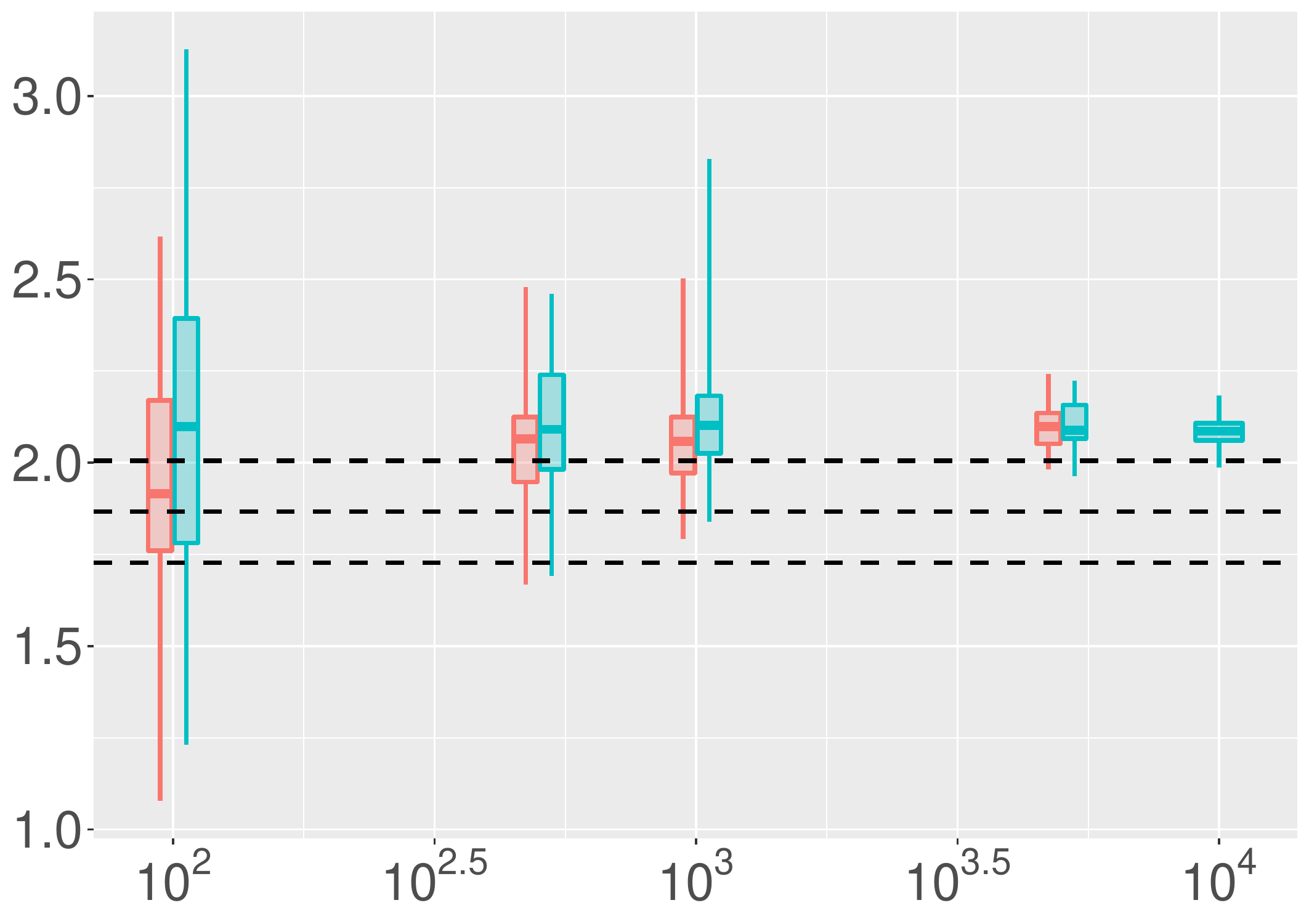}};
\node[below=of img2, xshift = -4cm, yshift = 1cm] (img5) {\includegraphics[width = 0.6\textwidth]{spatial_legend_bpf.pdf}};
\node[above=of img1, yshift = -1cm] {$(1, 1)$};
\node[above=of img2, yshift = -1cm] {$(2, 3)$};
\end{tikzpicture}
\caption{Filtering mean estimates obtained with DaC for node $(1, 1)$ and $(2, 3)$ of a $4\times4$ grid. The reference lines show the average value of the filtering mean estimate and the interquartile range obtained with 50 repetitions of a bootstrap PF with $N=10^5$ particles.
The boxplots from left to right report the distributions over 50 repetitions for $N=100, 500, 1000, 5000$ and $10000$. The results for the non-adaptive version of DaC and $N=10000$ are not included due to the excessive cost.}
\label{fig:spatial_app_cod}
\end{figure}
\fi
\end{document}